\documentclass[fleqn,usenatbib]{mnras}

\usepackage{newtxtext,newtxmath}

\usepackage[T1]{fontenc}

\DeclareRobustCommand{\VAN}[3]{#2}
\let\VANthebibliography\thebibliography
\def\thebibliography{\DeclareRobustCommand{\VAN}[3]{##3}\VANthebibliography}

\usepackage{graphicx}	
\usepackage{amsmath}	
\usepackage{amssymb}	
\usepackage[capitalise]{cleveref}		
\usepackage{booktabs}		
\usepackage[dvipsnames]{xcolor}   
\usepackage{bm}				
\usepackage{soul}			
\numberwithin{equation}{section}	

\newcommand{\der}{\text{d}} 
\newcommand{\DBaker}{Baker et al. (in preparation)} 
\newcommand{\BRickett}{Rickett et al. (in preparation)} 

\title[The $\theta$-$\theta$ Diagram]{
The $\theta$-$\theta$ Diagram: Transforming pulsar scintillation spectra to coordinates on highly anisotropic interstellar scattering screens
}

\author[T. Sprenger et al.]{
Tim Sprenger,$^{1}$\thanks{E-mail: tsprenger@mpifr-bonn.mpg.de}
Olaf Wucknitz,$^{1}$
Robert Main,$^{1}$
Daniel Baker,$^{2,3}$
and Walter Brisken$^{4}$
\\
$^{1}$Max-Planck-Institut f\"ur Radioastronomie, Auf dem H\"ugel 69, 53121 Bonn, Germany
\\
$^{2}$Canadian Institute for Theoretical Astrophysics, University of Toronto, 60 St. George Street, Toronto, ON M5S 3H8, Canada \\
$^{3}$Department of Physics, University of Toronto, 60 St. George Street, Toronto, ON M5S 1A7, Canada\\
$^{4}$National Radio Astronomy Observatory, Soccoro, NM 87801, USA
}

\date{Accepted XXX. Received YYY; in original form ZZZ}

\pubyear{2020}

\begin{document}
\label{firstpage}
\pagerange{\pageref{firstpage}--\pageref{lastpage}}
\maketitle

\begin{abstract}

We introduce a novel analysis technique for pulsar secondary spectra. The power spectrum of pulsar scintillation (referred to as the ``secondary spectrum'') shows differential delays and Doppler shifts due to interference from multi-path propagation through the interstellar medium. We develop a transformation which maps these observables to angular coordinates on a single thin screen of phase-changing material. This transformation is possible without degeneracies in the case of a one-dimensional distribution of images on this screen, which is often a successful description of the phenomenon. The double parabolic features of secondary spectra are transformed into parallel linear features, whose properties we describe in detail. Furthermore, we introduce methods to measure the curvature parameter and the field amplitude distribution of images by applying them to observations of PSR B0834+06. Finally, we extend this formalism to two-dimensional distributions of images on the interstellar screen.

\end{abstract}




\section{Introduction}

Diffractive scintillation is observed in pulsar observations when the radiation is distributed over multiple paths through the interstellar medium (ISM) that interfere with each other at the observer. The resulting interference patterns -- typically quasi-periodic criss-cross patterns of the pulsar's brightness in the plane of time and frequency -- can only occur when the incoming radiation is spatially coherent. This is the case because pulsars are very compact sources.

Scintillation is sensitive to the geometry of propagation as well as its first-derivative temporal evolution. Additionally, it is formed by the properties of the medium the radiation is passing through. However, these effects of unknown complexity can be separated to first order from the geometry, which gives scintillation studies a huge potential in measuring the location and velocity of the pulsar and the lensing ISM structures -- a discipline which is known as scintillometry. Since this basically resembles single-dish interferometry, it is sensitive to extremely small structures in the ISM. Furthermore, if the propagation paths can be constrained, they can be used as baselines of an interferometer of interstellar dimensions.

As was first discovered by \citet{2001ApJ...549L..97S}, the power spectrum of the intensity of scintillating pulsars often shows parabolic structures consisting of a dominant arc and sometimes potential downward arclets of the same curvature. The canonical explanation of these structures is a thin screen of phase-changing material somewhere in the ISM between observer and pulsar; the optical properties of interstellar screens have been described by \citet{1998ApJ...505..928G}. The scintillation arc phenomenon has been successfully explained using this model by \citet{2004MNRAS.354...43W} and \citet{2006ApJ...637..346C}. Furthermore, \citet{2010ApJ...708..232B} found through VLBI imaging that the structures are indeed sourced by distinct paths of propagation that are mostly located on a line and thus represent a one-dimensional thin screen as forecasted by theoretical models. 

The nature of these compact structures in the ISM is still debated. \citet{2014MNRAS.442.3338P} and \citet{2018MNRAS.478..983S} propose corrugated reconnection sheets of plasma at the boundary between between magnetic field configurations in the ISM, whose surface density waves are the source of the varying electron density responsible for diffraction. \citet{2019MNRAS.486.2809G} and \citet{2019MNRAS.489.3692G} have built another model that proposes magnetic noodles of plasma which are stabilized by parallel magnetic field lines that force constant plasma density along them.

The power spectrum of the pulsar's dynamic spectrum -- the secondary spectrum -- can be understood as a power distribution over the signal's delay of arrival and its Doppler rate (frequency shift) for all paths of propagation relative to each other. However, the quantities of interest are the angles of the incoming paths of radiation. In an ideal case of these paths being confined to a one-dimensional line on the sky, there is a one-to-one translation of delay and Doppler rate with the pairs of angles of interfering paths of propagation. In this paper, we present a transformation which utilizes this correspondence to present pulsar secondary spectra in a more physical and easier to analyse space.

\begin{figure}
 \includegraphics[trim={0cm 0cm 0cm 0cm},clip,width=\columnwidth]{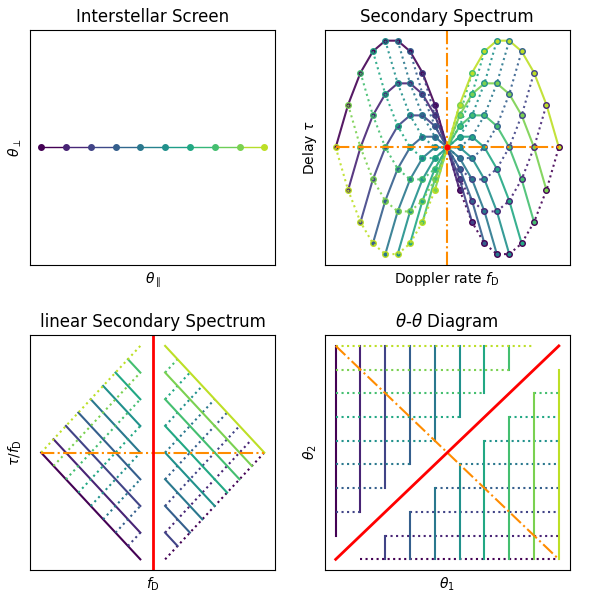}
 \vspace{-5mm}
 \caption{These sketches summarize the different data representations discussed in this paper. Features are colored in correspondence to the images on the screen that determine them. Since the secondary spectrum and its transformations arise from combinations of pairs of images, solid lines mark power belonging to the same first image while dashed lines mark power belonging to the same second image. This is also denoted in the markers shown in the secondary spectrum which are omitted in the lower panels. The red color marks a singularity in the transformations and the orange dotdashed line marks the typical location of increased noise and thus low sensitivity.
 }
 \label{fig:sketches}
\end{figure}

The upper panels of \cref{fig:sketches} show a scattering screen with marked propagation paths -- images -- and the parabolic structure they cause in the secondary spectrum. The lower panels of \cref{fig:sketches} then schematically introduce the transformations of these structure that will be discussed in the following sections. The transformation to a $\theta$-$\theta$ diagram is not only a good way to visualize the meaning of the secondary spectrum but also a tool to interpret this data. Straight lines are not only easier to interpret for the human eye but also for algorithms. For the latter point we refer the reader to \DBaker, where the $\theta$-$\theta$ diagram is used for precise curvature estimation and wavefield reconstruction. Here, the properties and meaning of the features visible in this data representation are described for reference, and guidance for their use and analysis is given. 

This paper is organized as follows: In Section \ref{sec:SecSpec_Definition} the formalism of secondary spectra is recapitulated, in Section \ref{sec:B0834_data} the data set used for example applications throughout this paper is described, in Section \ref{sec:nut_transform} the creation of secondary spectra by Fourier transforming with respect to $\nu\times t$ and $\nu$ instead of $t$ and $\nu$ is introduced, in Section \ref{sec:Obs_Abbr} the equations are abbreviated in terms of the observables, in Section \ref{sec:LinSecSpec_Curvature} we address the problem of measuring the curvature parameter, in Section \ref{sec:thth-diagram} the $\theta$-$\theta$ transformation is introduced and analysed for one-dimensional screens in Section \ref{sec:1D_analysis} and for two-dimensional screens in Section \ref{sec:2D_analysis}. We state our conclusions, and discuss future applications of our method in Section \ref{sec:conclusion}.

\section{The Secondary Spectrum in the Thin Screen Model}
\label{sec:SecSpec_Definition}

\begin{figure}
 \includegraphics[trim={0cm 1cm 0cm 0.5cm},clip,width=\columnwidth]{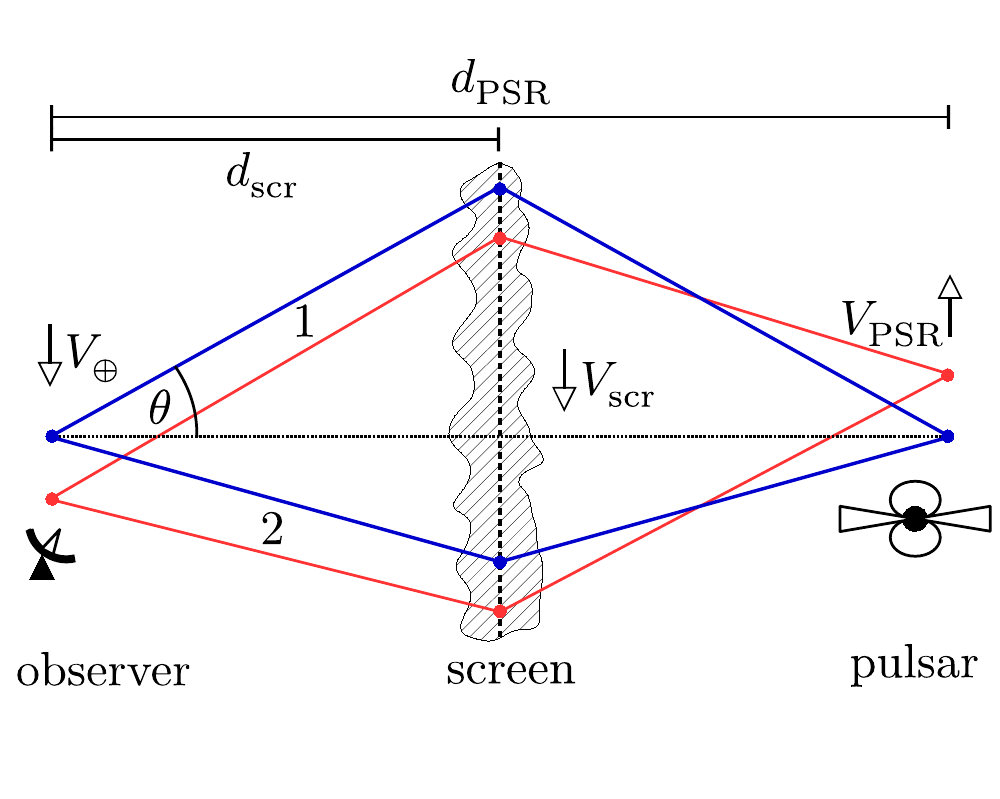}
 \vspace{-5mm}
 \caption{This sketch shows the physical model behind pulsar scintillation. The emission of the pulsar is focused onto the observer from multiple points (paths 1 and 2) of a thin screen of interstellar scattering material. All three parts of the system are moving, such that the optical paths evolve from the one shown in blue to the one shown in red after a certain time. Note that the horizontal scale is orders of magnitude larger than the vertical scale.
 }
 \label{fig:theory_def_sketch}
\end{figure}

\begin{figure}
 \centering
 \includegraphics[trim={0.5cm 0.5cm 0.5cm 0.5cm},clip,width=0.7\columnwidth]{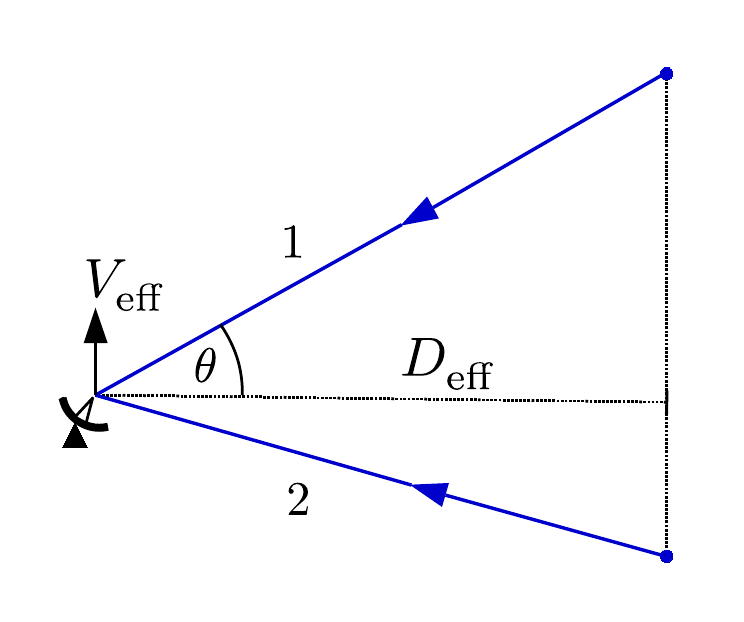}
 \vspace{-2mm}
 \caption{The physical model shown in \cref{fig:theory_def_sketch} can be substituted by a simpler effective picture where a collection of resting images at distance $D_\text{eff}$ radiate at the observer who is moving with the perpendicular velocity $V_\text{eff}$. The different path lengths cause the delay and the projections of the velocity on the directions of propagation cause the Doppler rate.
 }
 \label{fig:eff_def_sketch}
\end{figure}

In the thin screen model, the scattering screen is filled by a collection of images with two-dimensional angular positions $\bm{\theta}$. Although in principle all paths through the ISM screen contribute to the observed signal, rapidly varying phases cause destructive interference such that only paths of stationary phase remain \citep[see e.g.][]{1998ApJ...505..928G}. Images in the stationary phase approximation represent distinct paths of propagation of radiation emitted by the pulsar. As such, they differ from the ideal unperturbed emission by a complex factor whose magnitude constitutes an amplification and whose phase consists of an intrinsic phase shift induced by the refractive index of the scattering medium and a geometrical phase shift induced by the differing path length. Since it is a priori unknown if the signal results from resolved images or superimposed clusters of images, the most general parameterisation makes use of a two-dimensional amplitude field $B(\bm{\theta})$. The physical brightness can be identified with $B^2$, the square of the complex field amplitude $B$. This quantity absorbs the unknown intrinsic amplitude of the pulsar's emission into the amplification.

The dynamic spectrum is the integrated flux of full pulses as a function of time $t$ and frequency $\nu$. Thus, it is the square modulus of the electric field, which is determined by the sum of all images. In the thin screen model, it can thus be written in terms of phase shift differences of all pairs of images $(\bm{\theta}_1,\bm{\theta}_2)$:
\begin{align}
D(t,\nu) &\equiv \int \der^2\theta_1\der^2\theta_2 B(\bm{\theta}_1)B(\bm{\theta}_2) \\
&\hspace{0.8cm} \times \text{e}^{2\upi i\left[ \Delta\phi_\text{intr}(\bm{\theta}_1,\bm{\theta}_2,t,\nu) + \Delta\phi_\text{opt}(\bm{\theta}_1,\bm{\theta}_2,t,\nu) \right]} \, , \\
\Delta\phi_\text{opt} &= - \frac{\nu}{c}\left(\bm{\theta}_1-\bm{\theta}_2\right) \cdot \bm{V}_\text{eff}t \label{eq:delta_phi_opt_doppler} \\
&\hspace{0.4cm}+ \frac{\nu}{2c}D_\text{eff}\left(\bm{\theta}_1^2-\bm{\theta}_2^2\right) \, . \label{eq:delta_phi_opt_delay}
\end{align}
Above, $c$ is the speed of light and $D_\text{eff}$ and $\bm{V}_\text{eff}$ are the effective distance and the effective velocity, respectively. The effective quantities are defined by the distances of pulsar and screen and by velocities of earth, screen and pulsar perpendicular to the line of sight:
\begin{align}
    D_\text{eff} &\equiv \frac{d_\text{scr} d_\text{psr}}{d_\text{psr}-d_\text{scr}} \, , \\
    \bm{V}_\text{eff} &\equiv -\frac{d_\text{psr}}{d_\text{psr}-d_\text{scr}}\bm{V}_\text{scr} + \bm{V}_\oplus + \frac{d_\text{scr}}{d_\text{psr}-d_\text{scr}}\bm{V}_\text{psr} \, .
\end{align}
The physical model described here is visualized in \cref{fig:theory_def_sketch}. The time-independent part \cref{eq:delta_phi_opt_delay} of the optical phase difference is caused by the delayed pulse arrival due to different path lengths. Since the pulsar, ISM as well as the earth are moving, these path lengths change, which to first order results in \cref{eq:delta_phi_opt_doppler}, a relation proportional to time. As such, this relation is equal to the time derivative of the delay, and is therefore proportional to $\theta$ instead of $\theta^2$.

Condensing all geometric quantities into the effective distance and velocity corresponds to substituting the model by a screen filled with images at the effective distance moving with the effective velocity as shown in \cref{fig:eff_def_sketch}. This picture explains why the time-dependent term can be understood as a differential Doppler shift, namely the Doppler rate.

The zero-point on the screen is chosen to be the intersection with the line of sight at the start of the observation. If there is a dominant straight line of images on the screen, we can use this line as the x-axis and define positions on the screen by their distance $\theta$ to the centre and their angle $\alpha$ with respect to the positive side of this axis:
\begin{align}
    \bm{\theta} &\equiv \theta \left(\cos \alpha,\sin \alpha\right)^\intercal \, , \\
    \bm{V}_{\text{eff}} &\equiv V_\text{eff} \left(\cos \beta,\sin \beta\right)^\intercal \, .
\end{align}
Here, the effective velocity vector $\bm{V}_{\text{eff}}$ has been expressed by its norm $V_\text{eff}$ and its angle $\beta$ relative to the screen's axis.

The time and frequency dependence of the intrinsic phase is usually neglected on the scales of single observations.
Thus, the secondary spectrum is a map of optical phase differences.
The Fourier conjugates are physically interpreted as Doppler rate $f_\text{D}$ with respect to time and delay $\tau$ with respect to frequency:
\begin{align}
 \tau &= \frac{1}{2c}D_\text{eff}(\theta_1^2-\theta_2^2) \, , \label{eq:tau_def} \\
 f_\text{D} &\simeq - \frac{\nu}{c}V_{\text{eff}} \left[\theta_1 (\cos\beta\cos\alpha_1+\sin\beta\sin\alpha_1)\right. \nonumber \\
 &~\left. -\theta_2 (\cos\beta\cos\alpha_2+\sin\beta\sin\alpha_2) \right]  \, , \label{eq:fD_def}
\end{align}
where $\nu_0$ is the centre frequency of the band. Here we implicitly assume that the bandwidth of the dynamic spectrum is small compared to its center frequency. The generalization is discussed in Section \ref{sec:nut_transform}. We will now consider the case of all images lying strictly on a one-dimensional line, such that $\alpha=0$ or $\upi$. Hence, we treat $\theta$ as a signed number and remove $\alpha$ from the equations:
\begin{align}
    f_\text{D} &\simeq - \frac{\nu_0}{c}V_{\text{eff}} (\theta_1-\theta_2) \cos\beta  \, .
\end{align}
It is beneficial to define the secondary spectrum $S$ as the complex Fourier transform rather than the power spectrum as it is often done. In summary, the secondary spectrum connects to the dynamic spectrum as
\begin{align}
    \left|S\right|(f_\text{D},\tau) &\equiv \left| \int\der t\,\der\nu\, D(t,\nu)\,\text{e}^{-2\upi i[f_\text{D}t+\tau\nu]} \right| \label{eq:def_secspec_integral} \\
    &\propto \int \der^2\theta_1\der^2\theta_2\, B(\bm{\theta}_1)B(\bm{\theta}_2)~~\times \nonumber \\
    &\hspace{.5cm}\delta(f_D-f_D(\bm{\theta}_1,\bm{\theta}_2))\delta(\tau-\tau(\bm{\theta}_1,\bm{\theta}_2)) \, . \label{eq:secspec_to_deltas}
\end{align}
Although Dirac delta functions are used in \cref{eq:secspec_to_deltas}, the secondary spectrum cannot be understood as a field of infinite resolution in $(f_\text{D},\tau)$. Being a Fourier transform, its resolution is determined by the bandwidth and timespan of the dynamic spectrum. These are limited by observational restrictions as well as theoretically; a model of a screen that is static beyond common motion and has the same phase geometry for all frequencies is an approximation that must break down at some point.

Using the amplitude of the Fourier transform instead of the power spectrum allows us to utilize the correspondence $|S|\propto B(\bm{\theta}_1)B(\bm{\theta}_2)$ and prevents the distortion of the behaviour of noise.

\section{Example Data Set: PSR B0834+06}
\label{sec:B0834_data}

To visualize the techniques and transformations described in the following sections, we use data of the pulsar B0834+06 as an example here.
Over the last two decades, this pulsar has become one of the best examples of strong scintillation. The data we use were taken on 2005 November 12 by \citet{2010ApJ...708..232B} at the Arecibo observatory. Although this data set also includes VLBI measurements which have been successfully used to further constrain the interstellar screen, we will use single-dish data for demonstration. It consists of the pulsar's intensity and ranges over 6815 seconds (5700\,s on-source) and 32\,MHz (310.5-342.5\,MHz), sampled in 1364 time bins and 131072 frequency bins. For details of its reduction, we refer the reader to \citet{2019MNRAS.488.4952S}.

This data set has been shown by \citet{2010ApJ...708..232B,2019MNRAS.488.4963S} to agree very well with a one-dimensional thin screen. Only a particular feature at a delay of 1\,ms seems to originate from a more complicated geometry like an offset or a second screen, as studied in detail by \citet{2016MNRAS.458.1289L}. Other studies of this data set include \citet{2014MNRAS.442.3338P} who investigated a reconnection sheet model as the source of scintillation and an application of holography by \citet{2014MNRAS.440L..36P}.

The strong scintillation properties of pulsar B0834+06 have also been studied in other observations. To give an impression of the rich information available on the scintillation of this particular pulsar, we want to mention \citet{1997MNRAS.287..739R,2011AIPC.1357...97R,2001ApJ...549L..97S,2003ASPC..302..263S,2004MNRAS.354...43W,2008MNRAS.388.1214W,2005ApJ...619L.171H,2005MNRAS.362.1279W,2011ApJ...733...52G,2013MNRAS.429.2562T,2018MNRAS.480.4199F}.

In this manuscript, we use the values $V_\text{eff}=305$\,km/s and $\beta=-28.6^\circ$ when plotting quantities which cannot be measured directly using single-station data. These values approximately match the results from \citet{2010ApJ...708..232B} (see Table 4 therein).

\section{NuT Transform}
\label{sec:nut_transform}

The success of the analysis presented in this paper strongly depends on the sharpness of structures in the secondary spectrum. Thus, a good resolution in Doppler rate and delay is required while at the same time a good signal to noise ratio is needed. Therefore, being able to use the full available frequency range of the data to compute a single secondary spectrum is crucial. However, computing the power spectrum of a dynamic spectrum that ranges over a relatively wide band of frequencies leads to smearing effects caused by the frequency dependence of \cref{eq:fD_def}, which manifest in the spread of power belonging to the interference of one combination of images over a range of Doppler rates. These effects were quantitatively described by \citet{2019MNRAS.486.2809G}.

To correct for the smearing we scale the time information of each data point to a common frequency $\nu_0$ (the centre of the band):
\begin{align}
 \tilde{t} \equiv \frac{\nu}{\nu_0} t \, .
\end{align}
Thus, the Fourier transform is effectively performed with respect to $\nu t$, which we refer to as the \textit{NuT transform}. 

The Fourier transform with respect to $\tilde{t}$ is then performed as a direct Fourier transform. A much faster FFT is no longer possible because of the varying spacing between samples that was introduced by the scaling. A similar technique was applied to simulations in \citet{2019MNRAS.488.4963S}.

The NuT transform is not the only proposed solution to the problem of smearing effects in a standard FFT. Another method that led to successful sharpening of scintillation arcs is to perform the Fourier transform with respect to wavelength $\lambda$ instead of frequency \citep[see e.g.][]{2014JGRA..11910544F}. To compare these methods it is worth looking at the phase part of the Fourier integral \cref{eq:def_secspec_integral}:
\begin{align}
    \Phi &= - \frac{\nu t}{c}\left(\bm{\theta}_1-\bm{\theta}_2\right) \cdot \bm{V}_\text{eff} + \frac{\nu}{2c}D_\text{eff}\left(\bm{\theta}_1^2-\bm{\theta}_2^2\right) -f_\text{D}\tilde{t} - \tau \tilde{\nu} \\
    &\equiv \hat{f}_\text{D}\frac{\nu t}{\nu_0} + \hat{\tau}\nu -f_\text{D}\tilde{t} - \tau \tilde{\nu} \, .
\end{align}
Here the integration variables were replaced by $\tilde{t}$ and $\tilde{\nu}$ that differ for each method and the phase was reformulated in terms of the desired outcomes $\hat{f}_\text{D}$ and $\hat{\tau}$. Non-vanishing contributions to the secondary spectrum arise if $\der \Phi / \der t = \der \Phi / \der \nu = 0$. The resulting Fourier modes from this condition are shown in \cref{tab:De-smearing}.

\begin{table}
\begin{center}
\caption{This table shows a comparison of de-smearing methods (FFT: standard, NuT: time scaled by frequency, $\lambda$: wavelength instead of frequency) that differ by the parameters $\tilde{t}$ and $\tilde{\nu}$ used to perform the Fourier transform of the dynamic spectrum.
}
\label{tab:De-smearing}
\begin{tabular}{c|cccc} 
 \toprule 
method & $\tilde{t}$ & $\tilde{\nu}$ & $f_\text{D}$ & $\tau$ \\
\midrule 
FFT & $t$ & $\nu$ & $\frac{\nu}{\nu_0}\hat{f}_\text{D}$ & $\hat{\tau} + \frac{t}{\nu_0}\hat{f}_\text{D}$ \\
NuT & $\frac{\nu}{\nu_0}t$ & $\nu$ & $\hat{f}_\text{D}$ & $\hat{\tau}$ \\
$\lambda$ & $t$ & $\frac{c}{\nu}$ & $\frac{\nu}{\nu_0}\hat{f}_\text{D}$ & $-\frac{\nu^2}{c}(\hat{\tau}+\frac{t}{\nu_0}\hat{f}_\text{D})$ \\
\bottomrule 
\end{tabular}
\end{center}
\end{table}

As mentioned above, the NuT transform asserts the desired results, while the standard FFT approach produces a dependence on $\nu$ and hence a smearing for larger bandwidths. The term $\frac{t}{\nu_0}\hat{f}_\text{D}$ is subdominant in the cases regarded here. Note that the Fourier transform with respect to $\lambda$ still produces smearing for larger bandwidths and thus is not a good choice for applying to data with inverted arclets, or other discrete features in the secondary spectrum, as are in our example data set.
However, this transformation ensures $\tau/f_\text{D}^2 = -\frac{1}{c}\,\hat{\tau}/\hat{f}_\text{D}^2$. Therefore, this method successfully prevents smearing of the main arc, which is sufficient in cases of weak scintillation.

\citet{2016ApJ...818..166L} propose yet another technique of rescaling the axes before performing the Fourier transform, in order to preserve the characteristic sizes of scintles over large ranges of frequency. This method is motivated by Kolmogorov turbulence which is incompatible with the assumption of frequency-independent image locations made here. Since the following analysis relies on this assumption, this method cannot be used here while we have to focus on cases where this assumption is valid.

The NuT transform ensures that -- under the assumption of frequency-independent image positions -- the interference power of radiation arriving under the angles $\theta_1$ and $\theta_2$ is concentrated at the Doppler value of
\begin{align}
    f_\text{D} = - \frac{\nu_0}{c}V_{\text{eff}} (\theta_1-\theta_2) \cos\beta
\end{align}
under the assumption of a one-dimensional screen.

\begin{figure}
 \includegraphics[trim={1cm 0.5cm 3cm 1cm},clip,width=\columnwidth]{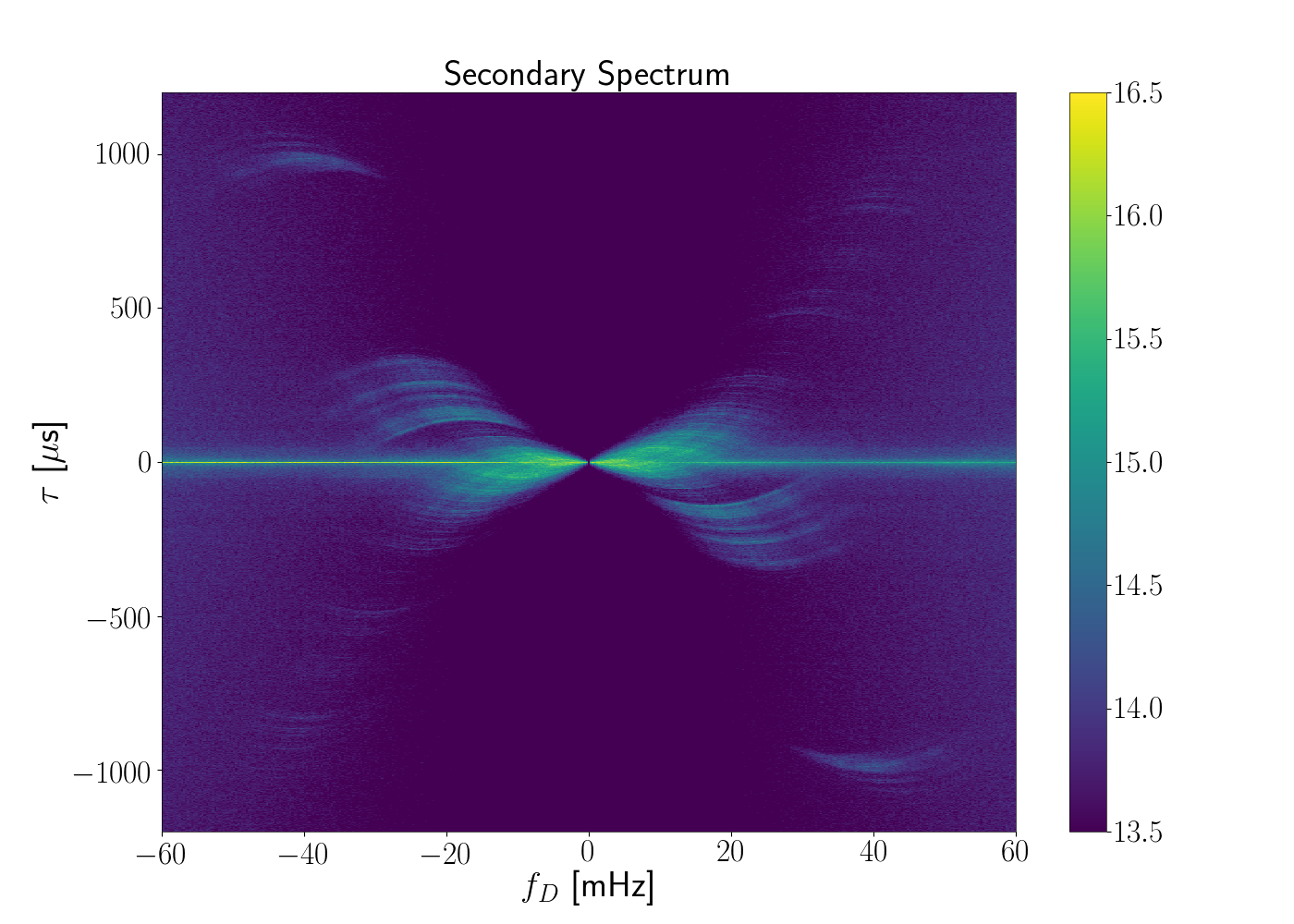}
 \vspace{-5mm}
 \caption{
 The Secondary Spectrum is multiplied by $|f_\text{D}|$ in order to get an equal weighting of angles on the screen. To reduce noise in this plot, each pixel is a sum of a square of 2 pixels width in $f_\text{D}$ and 20 pixels width in $\tau$. The power in arbitrary units is scaled logarithmically. The downward parabolic feature at $\tau = \pm 1$\,ms has received increased attention in the past because its apex is not located on the main parabola.
 }
 \label{fig:SecSpec_times_Doppler}
\end{figure}

A secondary spectrum of the example data set is shown in \cref{fig:SecSpec_times_Doppler}, which, in addition, has been rescaled according to the density of the ($\theta$,$\theta$)-space, as described in Section \ref{sec:thth-diagram}.

\section{Observables and Abbreviations}
\label{sec:Obs_Abbr}

Before discussing shapes and transformations of the power distribution in secondary spectra, it is beneficial to choose suitable variables to reduce degeneracy. These observables are sufficient to describe all of the information present in a secondary spectrum. Extending the common abbreviation of effective distance and velocity, we choose the following parameterisation:
\begin{align}
 \eta &\equiv \frac{c D_\text{eff}}{2\nu_0^2 V_{\text{eff}}^2\cos^2\beta} \, , \\
 \tilde{\theta}_i &\equiv - \frac{\nu_0 V_{\text{eff}}\cos\beta}{c}\theta_i \, , \\
 \gamma_i &\equiv  \frac{\cos(\beta-\alpha_i)}{\cos(\beta)} \, . \label{eq:def_gamma}
\end{align}
The curvature $\eta$ is the only global parameter. Beyond it, there are 4 parameters that represent the two-dimensional coordinates of the two interfering images. The parameter $\gamma=\cos(\alpha)+\tan(\beta)\sin(\alpha)$ is only present in a two-dimensional distribution of images. Otherwise it will always be equal to one. Using these parameters, \cref{eq:tau_def,eq:fD_def} reduce to
\begin{align}
 f_\text{D} &= \gamma_1\tilde{\theta}_1-\gamma_2\tilde{\theta}_2 \, , \label{eq:fD_start} \\
 \tau &= \eta(\tilde{\theta}_1^2-\tilde{\theta}_2^2) \label{eq:tau_start} \, .
\end{align}
Still, these parameters remain degenerate as long as there are one global and four specific parameters describing the position in a two-dimensional space. Hence, measuring the curvature is the first and most crucial step, discussed in Section \ref{sec:LinSecSpec_Curvature}. We will start by regarding the fully invertible case of a one-dimensional screen ($\gamma_i=1$). However, if we include the extra information of identifying structures that belong to the same image, we can also constrain a 2D screen, which will be discussed in Section \ref{sec:2D_analysis}.

\section{Linearization and Curvature Estimation}
\label{sec:LinSecSpec_Curvature}

As a global parameter, the curvature is not only a prerequisite to constrain the distribution of images, but it is the only information that is accessible if the power distribution is very diffuse. The estimation of the curvature can be difficult if the power is not distributed along a rather thin parabola but a more diffuse parabola or a convolution of a parabola with inverted parabolas, like it is the case in the interesting cases of strong scintillation where interference between subimages is significant.
In these cases, a Hough transform \citep[e.g.][]{2016ApJ...818...86B,2018MNRAS.476.5579X}, in which power is summed along thin parabolae of varying curvatures, is no longer reliable because of the multitude of overlaying structures present in the data.

If there is visible substructure, the parameter $\eta$ also manifests as the curvature of the downward arclets. In principle, this allows for multiple constraints on the curvature, which enables a much more precise estimation compared to probing the main arc alone. On top of that, we can transform parabolas into easier detectable linear features, as will be described below. We will discuss one method to use linear properties for curvature estimation below. Another method using the same information is shown by \DBaker~to lead to a very sensitive curvature estimation.

\begin{figure}
 \includegraphics[trim={1cm 0.5cm 3cm 1cm},clip,width=\columnwidth]{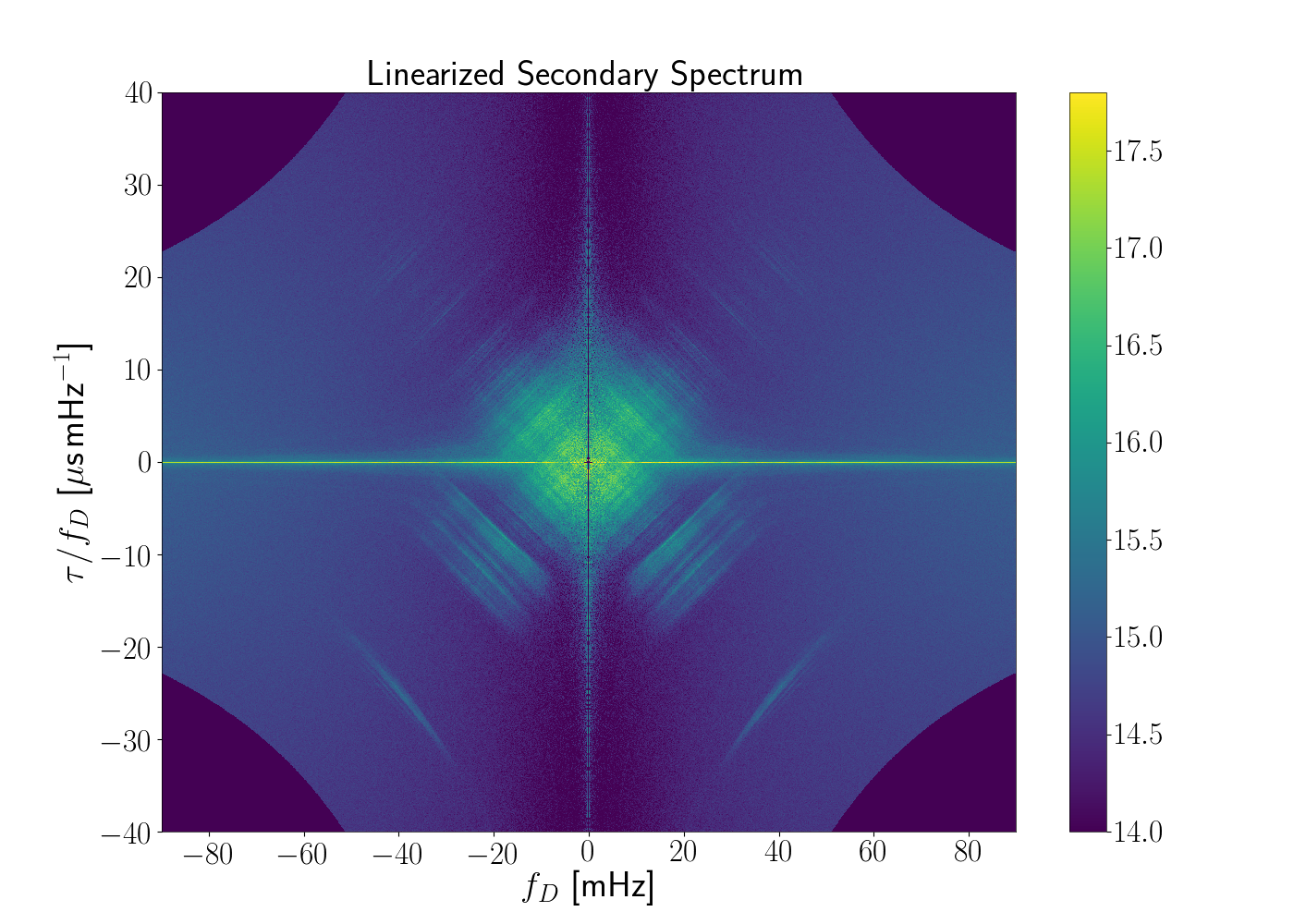}
 \vspace{-5mm}
 \caption{Linearized secondary spectrum created from the data shown in \cref{fig:SecSpec_times_Doppler}. The noise is not uniformly distributed because of the non-linearity of the transformation. The slope of the features is identical to the curvature $\eta$. The 1\,ms feature clearly has non-linear components.}
 \label{fig:linSecSpec}
\end{figure}

Before even measuring the curvature, the secondary spectrum can be linearized by transforming from $(f_\text{D},\tau)$-space to $(f_\text{D},\tau/f_\text{D})$-space, which results in expressions linear in $\theta$:
\begin{align}
 f_\text{D} &= \tilde{\theta}_1-\tilde{\theta}_2 \label{eq:lin_fd} \, , \\
 \frac{\tau}{f_\text{D}} &= \eta(\tilde{\theta}_1+\tilde{\theta}_2) \label{eq:tau_over_fd} \, .
\end{align}
This simple transformation has the advantage of transforming only one of the axes of the secondary spectrum. After choosing the parameter range and pixel width of the new $\tau/f_\text{D}$-axis, for each pixel we sum over the pixels in $\tau$ that belong to the width of a pixel in the new axis while accounting for the covered fraction of each pixel. This method is a bit slower than a simple one-to-one sampling but makes use of all available data and thus can lead to big improvements in signal to noise where the transformation increases the density of data. The result for the example data set is shown in \cref{fig:linSecSpec}.

We will now identify the parameter $\eta$ in the transformed data. By inserting \cref{eq:tau_over_fd,eq:lin_fd} into each other, the slope of lines of constant $\theta_1$ can be computed:
\begin{align}
 \frac{\tau}{f_\text{D}} &= \eta\left[\tilde{\theta}_1+\left(-f_\text{D}+\tilde{\theta}_1\right)\right] \\
 &= -\eta f_\text{D}+2\eta\tilde{\theta}_1 \, .
\end{align}
The corresponding result for constant $\theta_2$ is
\begin{align}
 \frac{\tau}{f_\text{D}} &= \eta\left[\left(f_\text{D}+\tilde{\theta}_2\right)+\tilde{\theta}_2\right] \\
 &= \eta f_\text{D}+2\eta\tilde{\theta}_2 \, .
\end{align}
Hence, slopes of features in this linearized secondary spectrum are a direct probe of the curvature $\eta$.

\begin{figure}
 \includegraphics[trim={0.5cm 0.5cm 3cm 1cm},clip,width=\columnwidth]{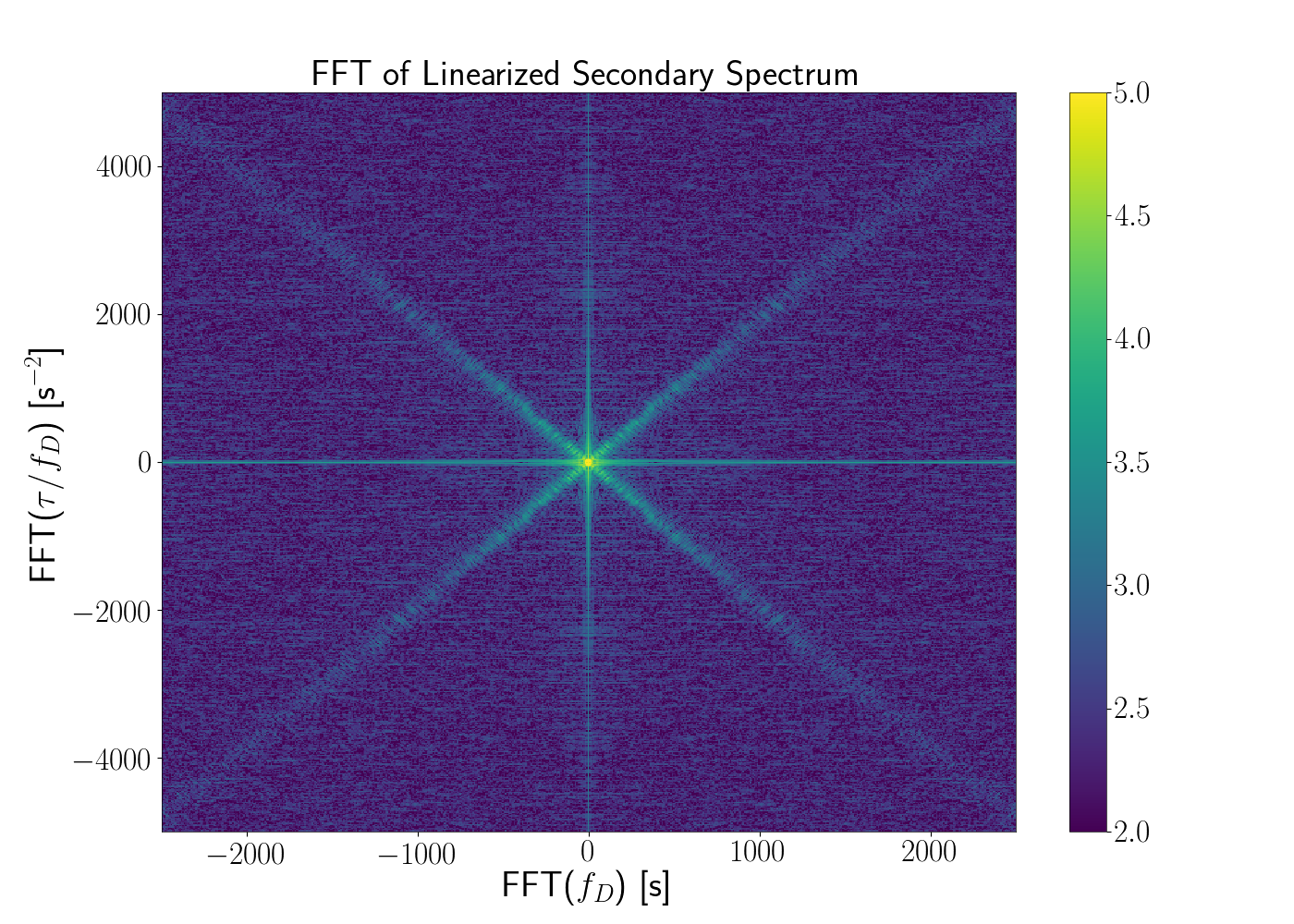}
 \vspace{-5mm}
 \caption{Amplitude of the Fast Fourier Transform of the logarithmic linearized secondary spectrum as shwon in \cref{fig:linSecSpec}. The scale and cutoff used exactly match \cref{fig:linSecSpec}. The slope of the feature in the positive quarter is $1/\eta$. The Fourier transform was computed from the interval ($\frac{\tau}{f_\text{D}}\in [-0.03,0.03]$\,s$^2$,$f_\text{D} \in [-0.06,0.06]$\,s).}
 \label{fig:FFT_lSS}
\end{figure}

\begin{figure}
 \includegraphics[trim={1,5cm 0.5cm 1cm 1cm},clip,width=\columnwidth]{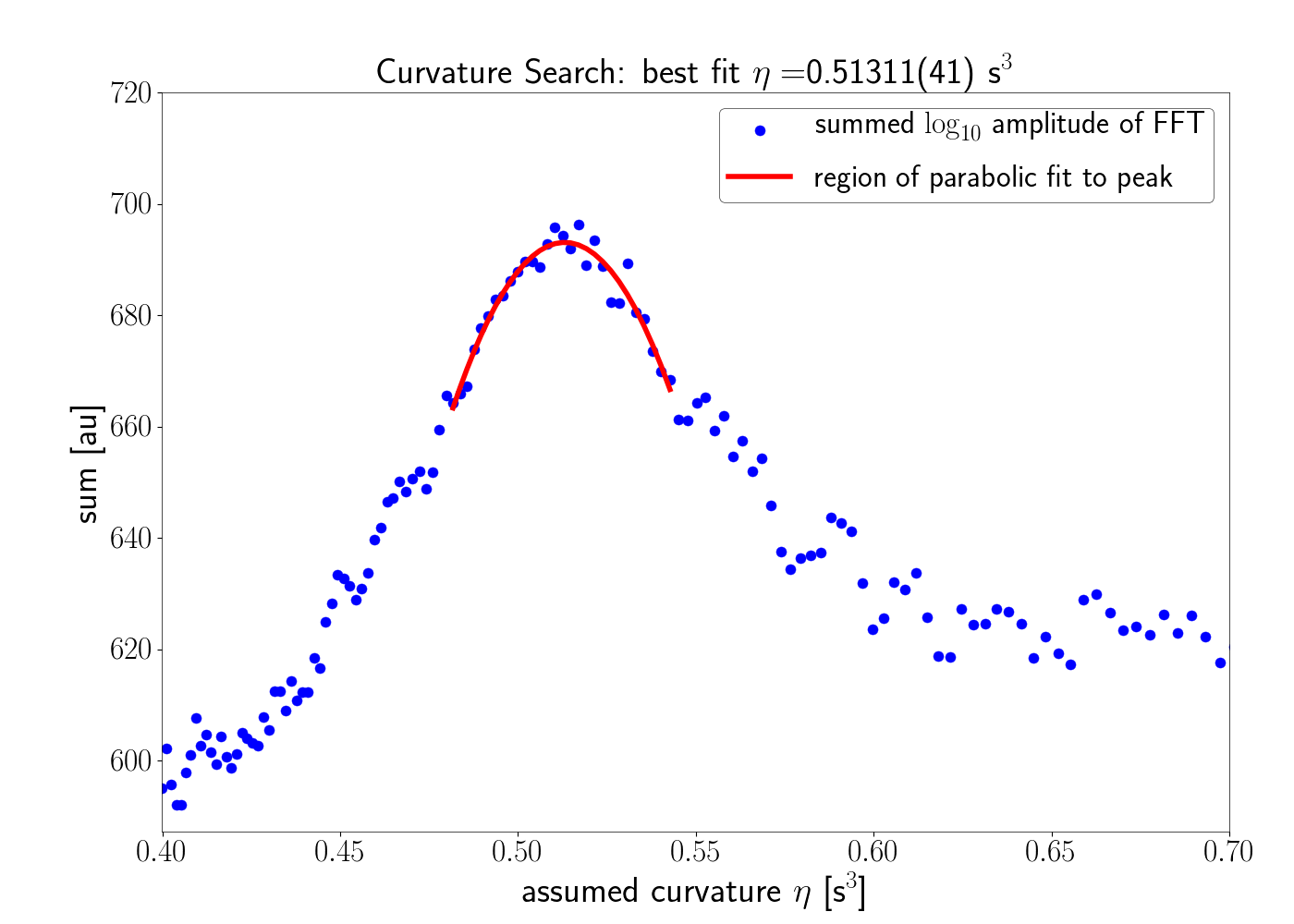}
 \vspace{-5mm}
 \caption{Sums over pixels of the Fast Fourier Transform (\cref{fig:FFT_lSS}) of the linearized secondary spectrum that lie on a line from the origin, whose slope is the reciprocal of the curvature $\eta$. The sum extends only over the upper right quarter and only over the inner part where the signal is visible by eye ($0<\text{FFT}(f_\text{D})\leq 2000$, $0<\text{FFT}(\tau/f_\text{D})\leq 5000$). Before summing, the decimal logarithm was applied to the absolute value of the pixels. The curvature was estimated by a parabolic fit to the peak.
 }
 \label{fig:FFT_lSS_sum}
\end{figure}

There are many possible ways to measure the slope of features. To prove the concept, we use the simple and fast option of performing a FFT. The slope of the lines results in preferred directions of Fourier modes, whose slope is the inverse curvature. To probe the visible features, the decimal logarithm is applied before performing the FFT, while all values below a certain threshold are dismissed. Since the linearized secondary spectrum is real and symmetric in $f_\text{D}$, the resulting power spectrum is highly symmetric such that the first quadrant contains the full information. The logarithmically weighted sum over linearly aranged pixels of Fourier amplitude in this quarter peaks at the correct curvature.
For the example data set, curvature estimation by this technique is shown in \cref{fig:FFT_lSS,fig:FFT_lSS_sum}. Here, the peak of the sum was determined by fitting a parabola. Note that the uncertainty of the fit potentially underestimates the full uncertainty. For a reference on the achievable accuracy of curvature measurements with this data set, see \DBaker.

\section{$\theta$-$\theta$ Transformation}
\label{sec:thth-diagram}

\begin{figure}
 \includegraphics[width=\columnwidth]{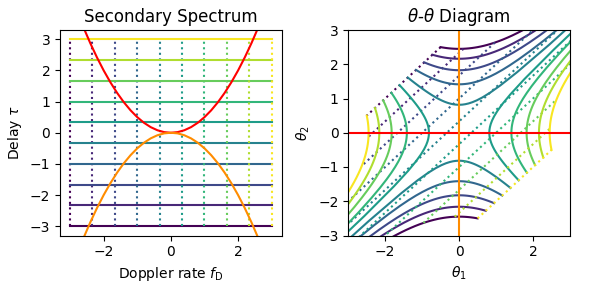}
 \vspace{-5mm}
 \caption{
 The grid of pixels representing the data present in the secondary spectrum (left panel), gets distorted under the $\theta$-$\theta$ transformation (right panel), which has to be taken into account not only for the position of samples taken from the former but also for their power which gets diluted or compressed. The red and orange lines mark the primary arc, where the signal usually strongest because of interference with the direct line of sight to the pulsar ($\theta_1=0$ or $\theta_2=0$).
 }
 \label{fig:sketch_pixels_SecSpec_to_thth}
\end{figure}

The motivation to transform the secondary spectrum from $(f_\text{D},\tau)$-space to $(\theta_1,\theta_2)$-space is to move from observational parameters to physical ones, which is possible because delay and Doppler rate resemble two constraints on two variables in the case of a perfectly one-dimensional screen. Since the power distribution is determined by a sum over all possible combinations of $\theta$ with itself, the $\theta$-$\theta$ diagram resembles the outer product of the field amplitude distribution vector with itself.

To derive the transformation, \cref{eq:lin_fd,eq:tau_start} are solved for $\theta_1$:
\begin{align}
 \tilde{\theta}_2 &= - f_\text{D} + \tilde{\theta}_1 \\
 \Rightarrow \tau &= \eta\left[ \tilde{\theta}_1^2 - \left( - f_\text{D} + \tilde{\theta}_1 \right)^2 \right] \\
 &= 2\eta f_\text{D} \left( -\frac{1}{2}f_\text{D} + \tilde{\theta}_1 \right) \\
 \Leftrightarrow \tilde{\theta}_1 &= \frac{1}{2}\left( \frac{1}{\eta}\frac{\tau}{f_\text{D}} + f_\text{D} \right) \, . \label{eq:th1_trafo}
\end{align}
The corresponding solution for $\theta_2$ is straightforward:
\begin{align}
 \tilde{\theta}_1 &=  f_\text{D} + \tilde{\theta}_2 \\
 \Rightarrow \tau &= \eta\left[ \left(  f_\text{D} + \tilde{\theta}_2 \right)^2 - \tilde{\theta}_2^2 \right] \\
 &= 2\eta f_\text{D} \left( \frac{1}{2}f_\text{D} + \tilde{\theta}_2 \right) \\
 \Leftrightarrow \tilde{\theta}_2 &= \frac{1}{2}\left( \frac{1}{\eta}\frac{\tau}{f_\text{D}} - f_\text{D} \right) \, . \label{eq:th2_trafo}
\end{align}
Since this transformation is non-linear, the area in $\theta$-$\theta$ space that is covered by a pixel in the secondary spectrum will change according to its location. This area $A_\theta$ can be computed by integrating over a single pixel, using the Jacobian determinant of the transformation:
\begin{align}
 \int \der \tilde{\theta}_1 \der \tilde{\theta}_2 &= \int \left|\frac{\partial (\tilde{\theta}_1,\tilde{\theta}_2)}{\partial (f_\text{D},\tau)}\right| \der f_\text{D} \der \tau  \\
 &\approx \left|\frac{\partial (\tilde{\theta}_1,\tilde{\theta}_2)}{\partial (f_\text{D},\tau)}\right| \delta f_\text{D} \delta \tau  \\
 &= \left|\frac{\partial \tilde{\theta}_1}{\partial f_\text{D}}\frac{\partial \tilde{\theta}_2}{\partial \tau} - \frac{\partial \tilde{\theta}_2}{\partial f_\text{D}}\frac{\partial \tilde{\theta}_1}{\partial \tau}\right| \delta f_\text{D} \delta \tau  \\
 &= \frac{1}{2\eta}\frac{1}{|f_\text{D}|}\delta f_\text{D} \delta \tau \, .
\end{align}
Hence, we can multiply the secondary spectrum by $|f_\text{D}|$ to get the interference power distribution of regions of equal size on the screen, as is shown in \cref{fig:SecSpec_times_Doppler}.

After inserting the scale factor from $\tilde{\theta}$ to $\theta$, we obtain an expression where the only physical parameter is the effective distance:
\begin{align}
    A_\theta = \int \der \theta_1 \der \theta_2 \simeq \frac{c\, \delta f_\text{D}\, \delta \tau}{D_\text{eff}|f_\text{D}|} \, .
\end{align}
When plotting the secondary spectrum after transforming it to ($\theta_1$,$\theta_2$)-space, the non-linearity of this transformation has several implications, as illustrated in \cref{fig:sketch_pixels_SecSpec_to_thth}. Firstly, the pixels from the secondary spectrum do not translate to uniformly shaped regions in the $\theta$-$\theta$ diagram, as was shown above. Since the Fast Fourier Transform (FFT) spaces pixels in such a way that more closely spaced Fourier modes would not be independent any more, this effectively means that a pixel in the secondary spectrum is an integration of the power over regions of varying size and shape in the $\theta$-$\theta$ diagram. Secondly, the density of pixel centres in the $\theta$-$\theta$ diagram can be overdense or underdense depending on the location and there is a divergence of the transformation at $f_\text{D}=0$.

Here, these issues are solved by the following strategy: For each uniformly sized and spaced pixel in the $\theta$-$\theta$ diagram, we compute the lowest and highest value for both $f_\text{D}$ and $\tau$. The corresponding area is then approximated as a square using the extremal values as border coordinates. Integrating over this square does not only give an amplitude corrected by density for the power, but also uses all available data which reduces the noise where possible. Before doing the integration, the bright centre point of the secondary spectrum is set to zero. Each pixel at the axes ($f_\text{D}=0$ or $\tau=0$) is replaced by the median value of the secondary spectrum, which is used as a noise approximation. However, instead of integrating over these pixels as done for the other pixels, their full value is added to the result whenever the integration covers them independent of the actual area covered. This last correction is chosen because this region is dominated by noise which should not get reduced by distributing it over smaller areas. Thus, this correction prevents unrealistically low results from this region of increased noise and of irregular structure.

In theory, the $\theta$-$\theta$ diagram shows the outer product of the field amplitude distribution $B(\theta)$ with itself, i.e.~the interference power of two paths approaching the observer under the angles $\theta_1$ and $\theta_2$. This results in horizontal and vertical lines that represent brighter regions in $\theta$. The field amplitude distribution vector could be read of by taking the square root of the diagonal where $\theta_1=\theta_2$. However, the diagonal represents the centre point of the secondary spectrum and thus contains no useful information. In addition to the noise, which is especially strong on the other diagonal due to gaps in the data caused by switches to a calibrator source and pulse to pulse intensity variations, the correct transformation parameters have to be identified before the diagram can be analysed.

\begin{figure}
 \includegraphics[trim={1.5cm 0.5cm 3cm 1cm},clip,width=\columnwidth]{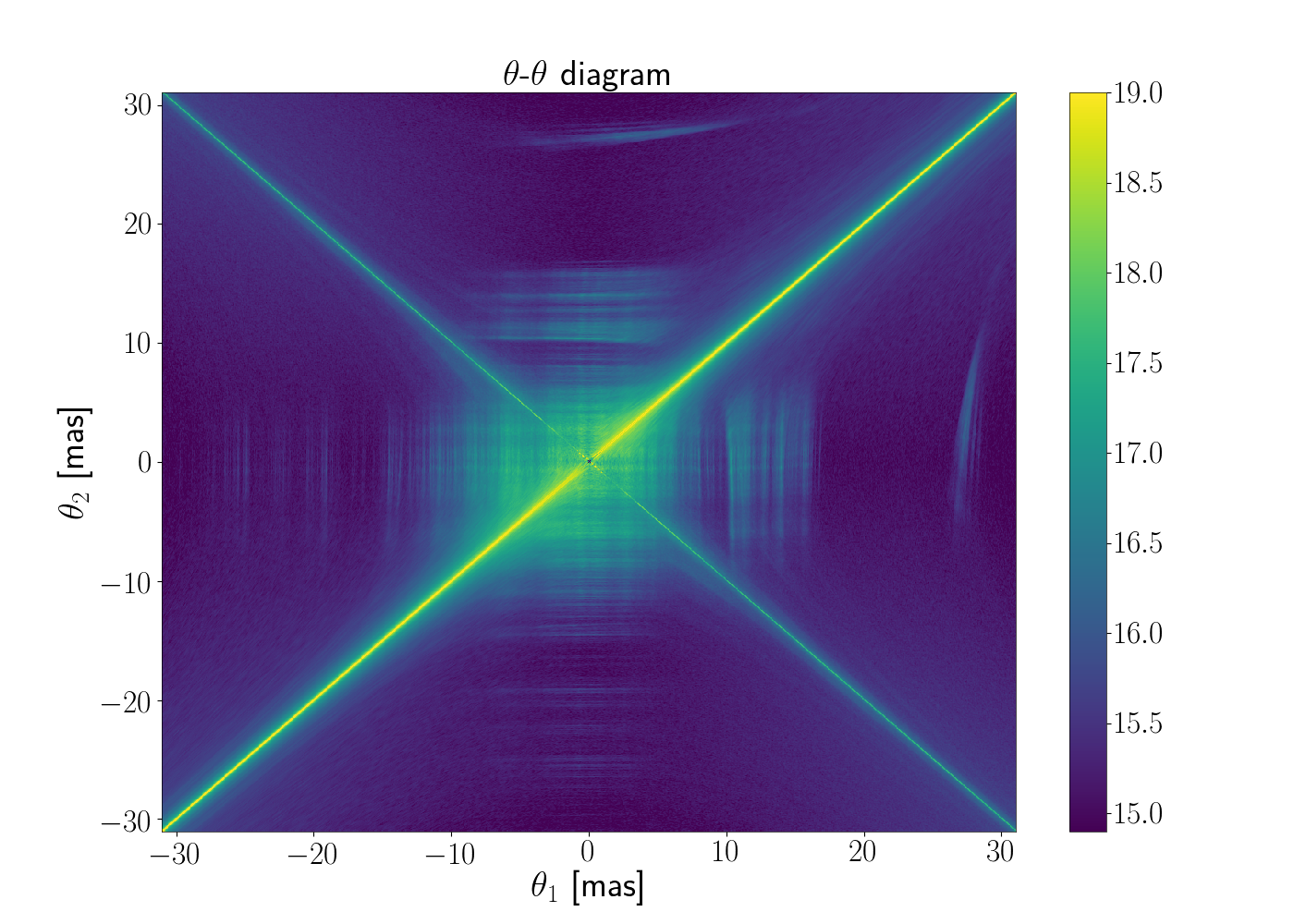}
 \vspace{-5mm}
 \caption{
 $\theta$-$\theta$ diagram computed using a good estimate for the curvature ($\eta=0.513$\,s$^3$ at $\nu=326.5$\,MHz). The power in arbitrary units is scaled logarithmically. The parallel linear slope of all features but the 1\,ms feature and the diagonals is clearly visible. The decreased signal to noise in the corners gives a good estimate of how bright a feature had to be to be visible there.
 }
 \label{fig:thth-diagram_good}
\end{figure}

\begin{figure}
 \includegraphics[trim={1.5cm 0.5cm 3cm 1cm},clip,width=\columnwidth]{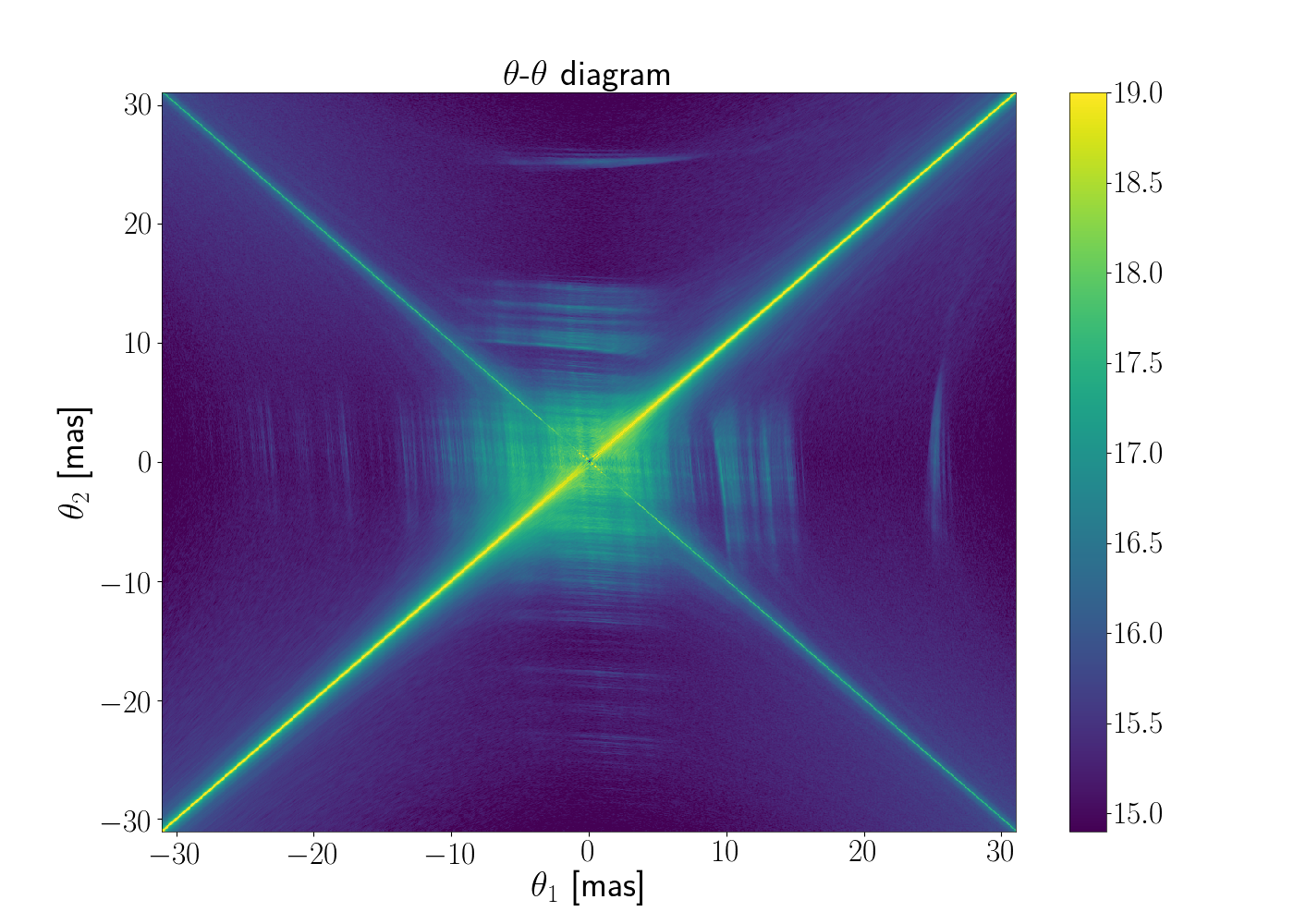}
 \vspace{-5mm}
 \caption{
 A $\theta$-$\theta$ diagram using a bad estimate for the curvature ($\eta=0.6$\,s$^3$ at $\nu=326.5$\,MHz). The power in arbitrary units is scaled logarithmically. The lines are not parallel to the axes anymore.}
 \label{fig:thth-diagram_bad}
\end{figure}

Except for the scaling of the axes, which is controlled by the effective velocity vector, the $\theta$-$\theta$ diagram only depends on the curvature. If a wrong curvature is chosen to transform the secondary spectrum, the structure in the $\theta$-$\theta$ diagram gets tilted such that interferences involving the same angle are no longer located at that angle, resulting in a visible shear of all structures in the diagram.
When the curvature $\eta'$ is used for the transformation, the slope of these lines can be computed. Using the notation $\varepsilon \equiv \eta/\eta'$, we insert the correct dependencies of delay and Doppler rate -- \cref{eq:lin_fd,eq:tau_start} -- into the transformation to $\theta_1$:
\begin{align}
 \tilde{\theta}_1' &\equiv \frac{1}{2}\left(\frac{\tau}{\eta' f_\text{D}} + f_\text{D} \right) \\
 &= \frac{1}{2}\left(\frac{\eta(\tilde{\theta}_1^2-\tilde{\theta}_2^2)}{\eta'  (\tilde{\theta}_1-\tilde{\theta}_2)} + \tilde{\theta}_1-\tilde{\theta}_2 \right) \\
 &= \frac{1}{2}\left(\varepsilon(\tilde{\theta}_1+\tilde{\theta}_2) + \tilde{\theta}_1-\tilde{\theta}_2 \right) \\
 &= \frac{\varepsilon+1}{2}\tilde{\theta}_1+\frac{\varepsilon-1}{2}\tilde{\theta}_2 \, .
\end{align}
The same logic applied to $\theta_2$ yields
\begin{align}
 \tilde{\theta}_2' &\equiv \frac{1}{2}\left(\frac{\tau}{\eta' f_\text{D}} - f_\text{D} \right) \\
 &= \frac{1}{2}\left(\frac{\eta(\tilde{\theta}_1^2-\tilde{\theta}_2^2)}{\eta'  (\tilde{\theta}_1-\tilde{\theta}_2)} - (\tilde{\theta}_1-\tilde{\theta}_2) \right) \\
 &= \frac{1}{2}\left(\varepsilon(\tilde{\theta}_1+\tilde{\theta}_2) - \tilde{\theta}_1+\tilde{\theta}_2 \right) \\
 &= \frac{\varepsilon-1}{2}\tilde{\theta}_1+\frac{\varepsilon+1}{2}\tilde{\theta}_2 \, .
\end{align}
For constant values of $\theta_1$, we can now insert the two results into each other such that we obtain $\theta_1'$ as a function of $\theta_2'$:
\begin{align}
 \tilde{\theta}_1' &= \frac{\varepsilon+1}{2}\tilde{\theta}_1+\frac{\varepsilon-1}{2}\frac{2}{\varepsilon+1}\left(\tilde{\theta}_2'-\frac{\varepsilon-1}{2}\tilde{\theta}_1\right) \\
 &= \frac{\varepsilon-1}{\varepsilon+1}\tilde{\theta}_2'+\frac{1}{2}\left((\varepsilon+1)-\frac{(\varepsilon-1)^2}{\varepsilon+1}\right)\tilde{\theta}_1 \\
 &= \frac{\varepsilon-1}{\varepsilon+1}\tilde{\theta}_2' + \frac{2\varepsilon}{\varepsilon+1}\tilde{\theta}_1 \, .
\end{align}
The above result resembles one straight line for each $\theta_1$. If the curvature estimate was correct, i.e.~$\varepsilon=1$, then the result is a straight line parallel to the $\theta_2'$-axis, as expected. We can apply the same logic while keeping $\theta_2$ constant:
\begin{align}
 \tilde{\theta}_2' &= \frac{\varepsilon+1}{2}\tilde{\theta}_2+\frac{\varepsilon-1}{2}\frac{2}{\varepsilon+1}\left(\tilde{\theta}_1'-\frac{\varepsilon-1}{2}\tilde{\theta}_2\right) \\
 &= \frac{\varepsilon-1}{\varepsilon+1}\tilde{\theta}_1'+\frac{1}{2}\left((\varepsilon+1)-\frac{(\varepsilon-1)^2}{\varepsilon+1}\right)\tilde{\theta}_2 \\
 &= \frac{\varepsilon-1}{\varepsilon+1}\tilde{\theta}_1' + \frac{2\varepsilon}{\varepsilon+1}\tilde{\theta}_2 \, .
\end{align}
This result in turn resembles one straight line for each $\theta_2$ which turns for the correct curvature value into a line parallel to the $\theta_1'$-axis. Both results imply the same absolute value of slope with respect to the corresponding axis. Thus, we can obtain the correct curvature by measuring the slope $a\equiv (\varepsilon-1)/(\varepsilon+1)$:
\begin{align}
 a &= \frac{\eta - \eta'}{\eta+\eta'} \\
 \Leftrightarrow \eta &= \frac{1+a}{1-a}\eta' \, .
\end{align}
$\theta$-$\theta$ diagrams computed for the example data set using different curvature values are shown in \cref{fig:thth-diagram_good,fig:thth-diagram_bad}. For these diagrams, the secondary spectrum was subtracted by its median first to reduce noise, especially in the corners where many pixels are summed into few pixels.

The true curvature could be estimated by the same technique as in Section \ref{sec:LinSecSpec_Curvature}, however, it is more efficient to use the fact that the $\theta$-$\theta$ diagram only resembles an outer product if the correct curvature was used for the transformation. This knowledge is applied by \DBaker~in developing an efficient method to precisely measure arc curvatures.

Once a good curvature estimation is found, it can be used to compute a $\theta$-$\theta$ diagram for further analysis.

\section{One-dimensional Analysis}
\label{sec:1D_analysis}

If the scaling from ($f_\text{D}$,$\tau$)-space to ($\theta_1$,$\theta_2$)-space was correctly taken into account, the $\theta$-$\theta$ diagram $\Theta$ can be understood as the outer product of the field amplitude vector plus nonuniform noise:
\begin{align}
    \Theta(\tilde{\theta}_1,\tilde{\theta}_2) = B(\tilde{\theta}_1) B(\tilde{\theta}_2) + N(\tilde{\theta}_1,\tilde{\theta}_2)\, .
\end{align}
This noise term has three contributions: Firstly, there is background noise from the noise dominated parts of the secondary spectrum. This initially homogeneous noise becomes nonuniform through the $\theta$-$\theta$ transformation and therefore is most relevant in the corners. Secondly, there is additional noise close to the former $\tau=0$ axis, located around the $\theta_1=-\theta_2$ axis. Lastly, there is noise due to signal leaking from one pixel into another which is caused by the finite sampling size in the secondary spectrum. Because of the transformation, the last source of noise is especially prominent on the $\theta_1=\theta_2$ line, representing the spread of the central bright pixel of the secondary spectrum. In summary, the $\theta$-$\theta$ diagram is dominated by noise on the diagonals and in the corners.

Physically, the noise contributions arise from a number of sources. On top of contamination from sources not belonging to the studied objects -- e.g.~the instrument and the sky -- observations of scintillation suffer from the source noise of the pulsar, i.e~variations of the pulsar's intensity that are not caused by propagation effects but are intrinsic. This is most visible in the pulse-to-pulse variations in time that cause the huge noise contribution along $\tau=0$, and is much larger than background noise. The NuT transform widens this noise distribution a bit in Doppler rate because the Fourier transform is no longer performed over the $t$ (see Section \ref{sec:nut_transform}). The same region is further contaminated by the Fourier transform of data gaps caused by observing a calibrator source. The $f_\text{D}=0$ region, that is irrelevant for $\theta$-$\theta$ diagrams, is contaminated by noise from Radio Frequency Interference (RFI) and band edge effects. Noise properties of B0834+06 were studied in detail by \cite{2011ApJ...733...52G} who also present a thorough theoretical analysis of the noise present in dynamic and secondary spectra for single-dish and interferometric observations, which is expanded to VLBI observations in \cite{2012ApJ...758....6G}.

\begin{figure*}
 \centering
 \includegraphics[trim={2cm 0cm 2cm 0cm},clip,width=.8\textwidth]{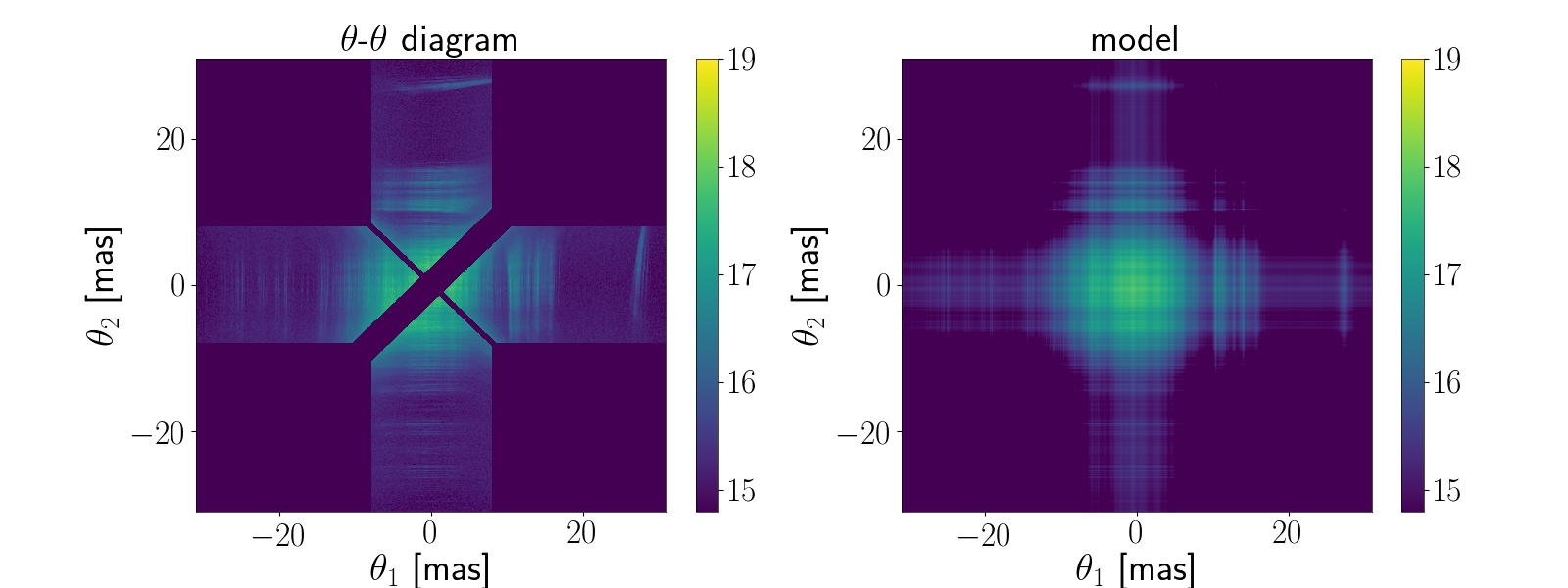}
 \vspace{-2mm}
 \caption{By masking the diagonals and the corners (left plot) the $\theta$-$\theta$ diagram is reduced to areas where the noise is subdominant. Hence, an eigenvector can be obtained by fitting its outer product as a model to the diagram. Note that the data and model are shown in logarithmic scales but the fit was carried out on the original linearly scaled data. As a result, artefacts appear more prominent than they are. The barycenter of power belonging to the 1\,ms feature (top of left plot) is clearly shifted away from the $\theta_1=0$ line. This behaviour cannot be captured by the eigenvector model, as it cannot be explained within a strictly one-dimensional model.
 }
 \label{fig:1D_fit}
\end{figure*}

Since the signal itself is very weak in the corners if the field amplitude vector has a shape peaked around $\theta=0$, this region can be masked without loss of information. If the vicinity of the diagonals is masked in addition, the noise becomes subdominant and a fit to the outer product of the field amplitude vector -- i.e.~the eigenvector of this matrix -- with itself can be attempted. An eigenvector decomposition (either of the complex matrix or only the amplitudes) relies on the knowledge of all matrix elements. If the diagonals and other critical regions have to be masked, the resulting dominant eigenvector will be distorted in a complicated way. For this paper we instead fit the vector directly by minimizing residuals of the good matrix elements. This has the advantage that masks do not introduce systematic errors. An example fit is shown in \cref{fig:1D_fit}. In addition, we could weight residuals according to the varying noise level across the matrix with this approach if a less aggressive mask than used for the example data is desired.

\section{Two-dimensional Analysis}
\label{sec:2D_analysis}

Although some features in \cref{fig:thth-diagram_good,fig:thth-diagram_good} do not resemble straight lines, contiguous curved lines can still be identified. A possible explanation is a still dominant one-dimensional structure of images to interfere with. Thus, we can approximately solve for a curve caused by an image $\bm{\theta_2}$ of two-dimensional coordinates interfering with a straight line $\bm{\theta_1}(\theta)$:
\begin{align}
    \bm{\theta_1} &\equiv (\theta,0)^\intercal \, ,\\
    \bm{\theta_2} &\equiv (\theta_2 \cos(\alpha),\theta_2 \sin(\alpha))^\intercal \, .
\end{align}
By using \cref{eq:th1_trafo,eq:th2_trafo}, we obtain the coordinates in the $\theta$-$\theta$ diagram as a function of one running parameter $\theta$:
\begin{align}
    \theta_1'(\theta) &= \frac{1}{2}\left(\frac{\theta^2-\theta_2^2}{\theta-\theta_2\gamma_2}+\theta-\theta_2\gamma_2\right) \\
    \theta_2'(\theta) &= \frac{1}{2}\left(\frac{\theta^2-\theta_2^2}{\theta-\theta_2\gamma_2}-\theta+\theta_2\gamma_2\right)
\end{align}
These lines can be identified in the diagram if the screen is sparse enough. However, the parameter $\gamma$ does not translate unambiguously to a location on the screen. If we consider positions on a full circle $-\upi\leq \alpha<\upi$, there are up to two solutions of \cref{eq:def_gamma}:
\begin{align}
    \gamma &= \frac{\cos(\beta-\alpha)}{\cos\beta} = \frac{\cos(\pm (\alpha-\beta) + 2\upi n)}{\cos\beta} \\
    \Leftrightarrow \alpha &= \beta \pm \arccos\left(\gamma\cos\beta\right)+2\upi n \, , \label{eq:alpha_of_gamma}
\end{align}
where $n \in \mathbb{Z}$.

\begin{figure}
 \includegraphics[trim={1.5cm 0.5cm 3cm 1cm},clip,width=\columnwidth]{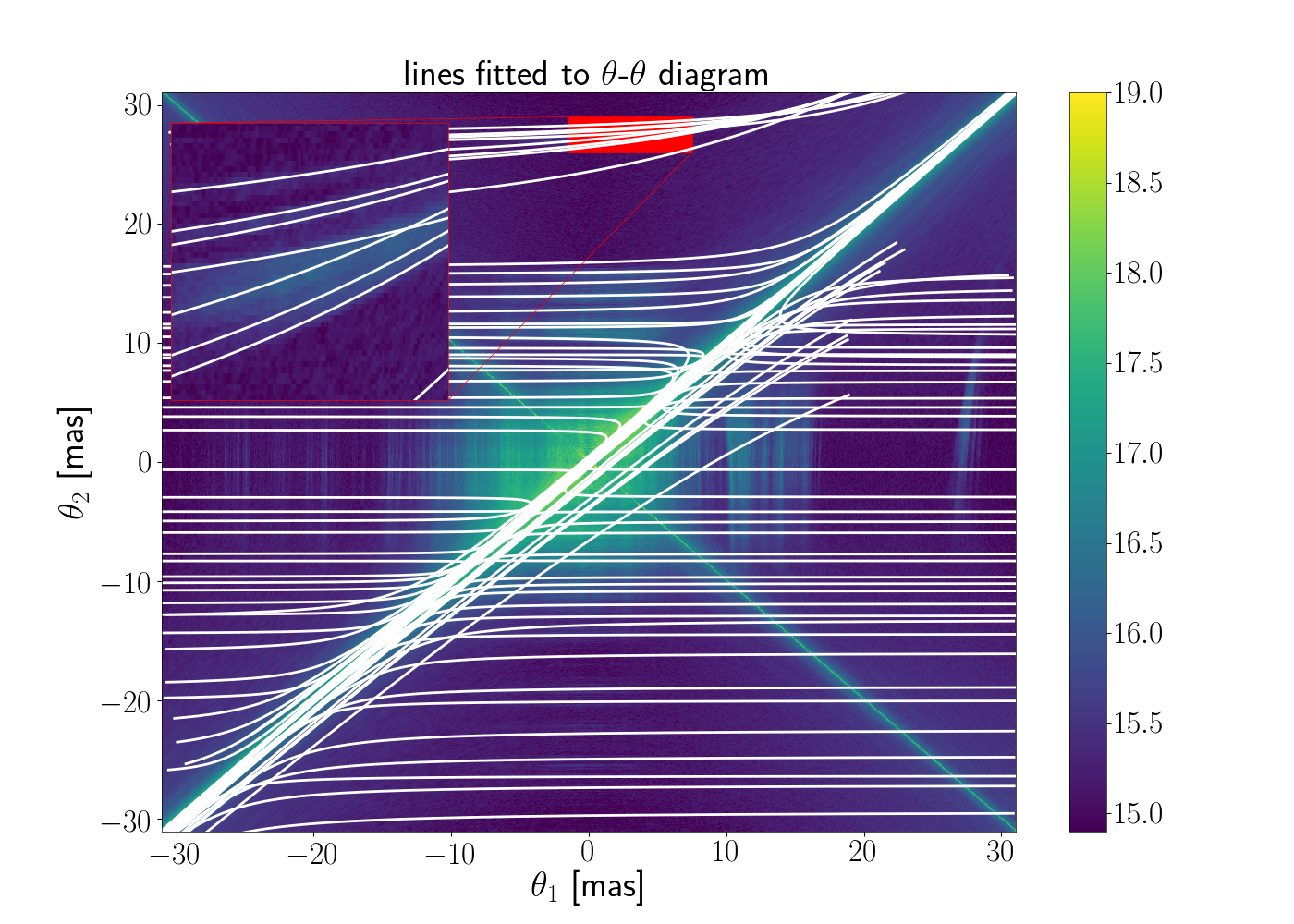}
 \vspace{-5mm}
 \caption{The lines shown above were fitted to features in the $\theta$-$\theta$ diagram that were identified by eye to belong to the same structure. The curvature of lines is most prominent near the $\theta_1=\theta_2$ diagonal. All lines diverge at this diagonal and reappear at the other side. In the case of the 1-ms feature this leads to structures that cross all other lines but have no imprint on the data because of their weakness far away from the centre. The region covered by a red rectangle contains the 1\,ms feature and is shown enlarged and rescaled in the inset. If this feature lies on the same screen, lines deviate from horizontal alignment corresponding to the deviation of their belonging image from the dominant one-dimensional distribution (compare \cref{fig:screen_solution}).
 }
 \label{fig:manual_lines}
\end{figure}

The identification of lines belonging to the same image is a complex task in the case of a two-dimensional screen and thus best done by eye. Here, the problem was solved by manually identifying a collection of features that belong to the same line. As shown above, these can then be fitted using the two parameters $\theta_2$ and $\gamma_2$. As a result, an irregularly spaced sample of images with ambiguous locations on the screen is obtained, shown in \cref{fig:manual_lines}.

\begin{figure*}
 \centering
 \includegraphics[trim={0cm 0cm 0cm 0cm},clip,width=.9\textwidth]{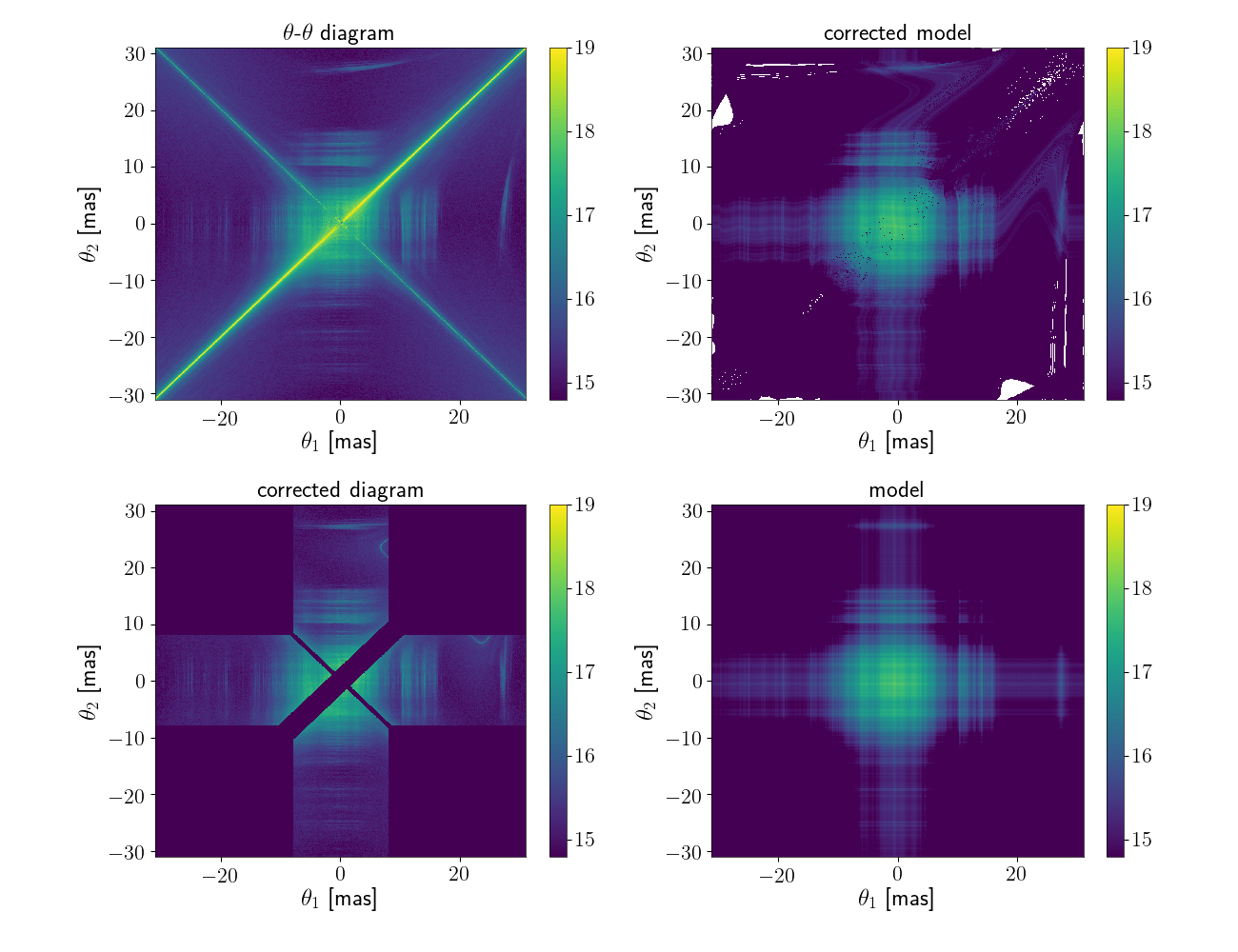}
 \vspace{-5mm}
 \caption{Upper left: Original $\theta$-$\theta$ diagram. Lower left: Applied correction for non-vanishing components in the second dimension and applied mask. Lower right: Result of the fit. Upper right: Transformation of the fit result using the two-dimensional information. The technique successfully reproduces the 1-ms feature and some wobbling in the lines, but fails at overlapping structures (especially close to the origin).
 }
 \label{fig:2D_fit}
\end{figure*}

If the screen is sparse enough such that there exists only one image for the same value of $\theta_2$, we can interpolate this sample in order to get a contiguous function $\gamma_2(\theta_2)$. Effectively, the two-dimensional screen is now replaced by a one-dimensional field amplitude distribution $B(\theta)$ and a distorting function $\gamma(\theta)$. Thus, this distortion can be removed by sampling $(\theta_1',\theta_2')$ as a function of $(\theta_1,\theta_2)$ following
\begin{align}
    \theta_1'(\theta_1,\theta_2) &= \frac{1}{2}\left(\frac{\theta_1^2-\theta_2^2}{\theta_1\gamma(\theta_1)-\theta_2\gamma(\theta_2)}+\theta_1\gamma(\theta_1)-\theta_2\gamma(\theta_2)\right) \\
    \theta_2'(\theta_1,\theta_2) &= \frac{1}{2}\left(\frac{\theta_1^2-\theta_2^2}{\theta_1\gamma(\theta_1)-\theta_2\gamma(\theta_2)}-\theta_1\gamma(\theta_1)+\theta_2\gamma(\theta_2)\right)
\end{align}
to get a corrected $\theta$-$\theta$ diagram. An example is shown in \cref{fig:2D_fit}, where the correction also was inverted on the fitted model. Visible problems of this approach are doubly counted pixels as well as pixels that cannot be assigned to a position in $(\theta_1,\theta_2)$-space at all.

\begin{figure}
 \centering
 \includegraphics[trim={4.5cm 0.5cm 3cm 1cm},clip,width=\columnwidth]{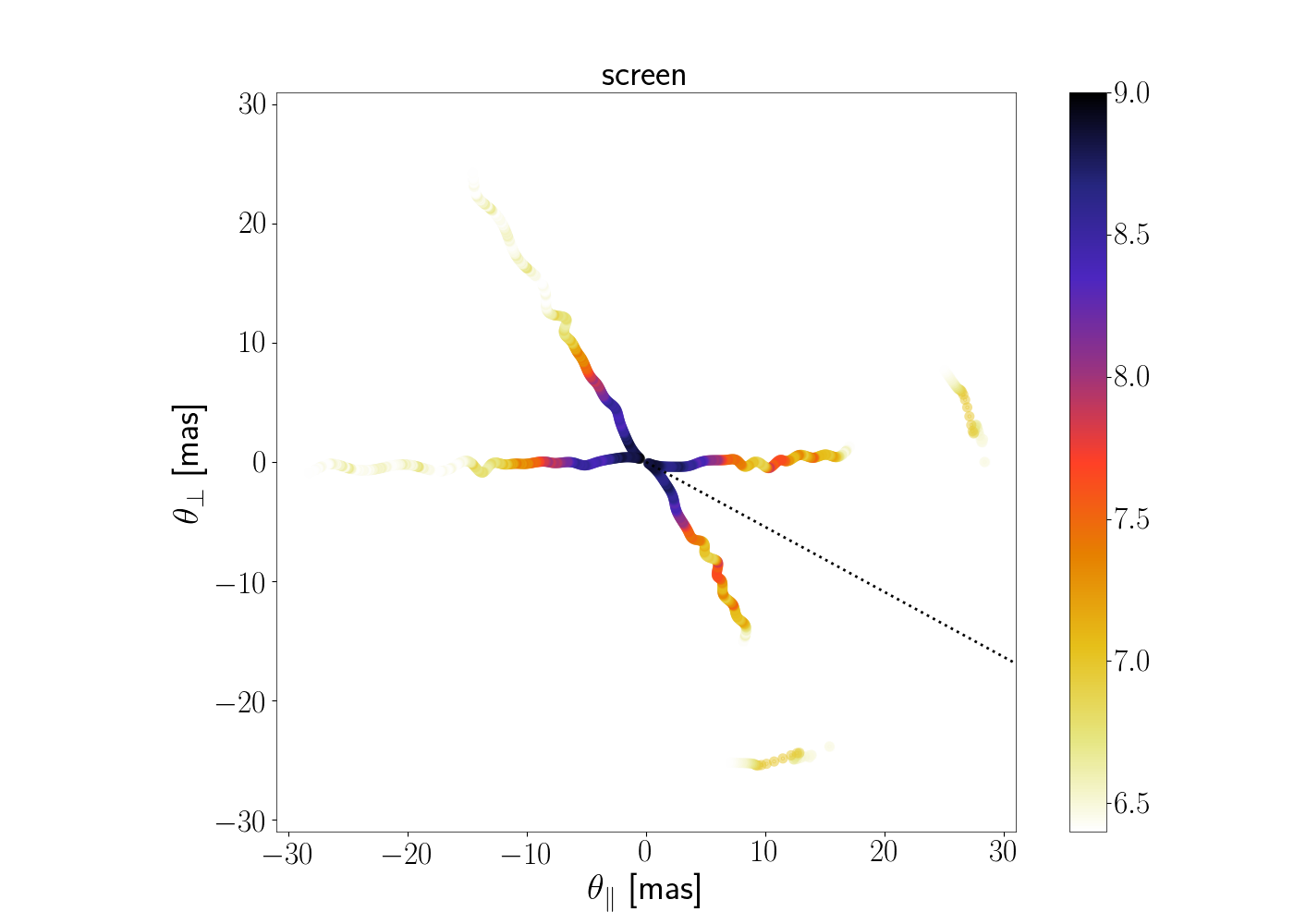}
 \vspace{-5mm}
 \caption{The two-dimensional screen model as inferred from single-dish data by interpolating manually identified lines suffers from an ambiguity that mirrors each point on the axis of the effective velocity, which is indicated by the black dotted line. The field amplitude distribution is shown in logarithmic scale. Since the dominant one-dimensional screen lies per definition at $\theta_\perp=0$, only the 1-ms feature is expected to deviate significantly from this axis such that its real position is the one at the bottom of the plot.
 }
 \label{fig:screen_solution}
\end{figure}

Using \cref{eq:alpha_of_gamma}, the fitted eigenvector can be translated onto the screen, as shown in \cref{fig:screen_solution}. As discussed above, this translation suffers from an ambiguity which produces two possible locations for most images. However, the lines were fitted using the assumption of a dominant one-dimensional distribution of images. Thus, most images and in particular the brightest ones have to lie at $\theta_\perp=0$.

\section{Conclusion}
\label{sec:conclusion}

This work introduces transformations of pulsar secondary spectra that convert parabolic features associated with discrete images on a scattering screen to linear features. In particular we introduce the $\theta$-$\theta$ diagram which translates a secondary spectrum caused by a one-dimensional screen to a data matrix that is equal to the outer product of the field amplitude distribution along these images.

We discuss the connection between the slopes of these linear features and the curvature of parabolas in the secondary spectrum, which is an important observable of scintillation. We introduce a method to utilize this finding for curvature estimation based on the properties of Fourier transforms.

By fitting for the eigenvector of the $\theta$-$\theta$ diagram, we obtain the field amplitude distribution of the one-dimensionally distributed images. This approach has the advantage of using the full available data set. We extend this analysis to two-dimensional distributions of images under the premise of these being sparse and dominated by a one-dimensional line of images. By correcting manually identified lines for their shape distortion due to deviations from the one-dimensional case, we can fit for the field amplitude distribution even in the two-dimensional case, up to an ambiguity creating up to two solutions for the location of every image.

We demonstrate all techniques introduced in this work on scintillation data of PSR B0834+06 taken by \citet{2010ApJ...708..232B}. This data set is chosen because of its high quality and resolution as well as for the occurrence of very strong scintillation that makes interference between images far from the central line of sight visible.

This work is focused on introducing new data transformations. Hence, we explain the morphology of observable features in detail, and introduce ways to make use of this transformation. We chose to apply them to a well studied data set in order to allow for easy comparison rather than presenting new results for a particular pulsar. For an application of the $\theta$-$\theta$ diagram to PSR B0450-18 we refer the reader to \BRickett.

As shown in \DBaker, it is possible to extend the concept of the $\theta$-$\theta$ diagram to include the complex phases of the secondary spectrum, which allows for even more precise constraints on the curvature.

The prevalence of thin and dominantly one-dimensional scattering screens in pulsar scintillation does not only offer good opportunities to study these ISM structures themselves but also makes them formidable tools to constrain geometric properties of the pulsars they are illuminated by. Directly employing their one-dimensional nature in techniques like the ones presented in this paper will further improve evolving applications of scintillometry like constraining orbital velocities, interstellar holography (resolving the ISM screen) and interstellar interferometry (using the ISM screen as an interferometer on astronomical scales to resolve the pulsar system). 

\section*{Acknowledgements}

We thank Dana Simard for sharing her results of data reduction performed on the example data set, as well as Vivek Venkatraman Krishnan and the anonymous referee for helpful comments.

Tim Sprenger is a member of the International Max Planck Research School for Astronomy and Astrophysics at the Universities of Bonn and Cologne. The Arecibo Observatory is operated by SRI International under a cooperative agreement with the National Science Foundation (AST-1100968), and in alliance with Ana G. \'{M}endez-Universidad Metropolitana, and the Universitie Space Research Association. The National Radio Astronomy Observatory is a facility of the National Science Foundation operated under cooperative agreement by Associated Universities, Inc.

\section*{Data Availability}


The interactive code used to manually fit lines to features in the $\theta$-$\theta$ diagram as shown in \cref{fig:manual_lines} is available at \href{https://github.com/SprengerT/scint_thth.git}{github.com/SprengerT/scint\_thth.git}. The exact parameters of the lines used for the analysis presented above are included.


\bibliographystyle{mnras}
\bibliography{thth}

\begin{thebibliography}{}
\makeatletter
\relax
\def\mn@urlcharsother{\let\do\@makeother \do\$\do\&\do\#\do\^\do\_\do\%\do\~}
\def\mn@doi{\begingroup\mn@urlcharsother \@ifnextchar [ {\mn@doi@}
  {\mn@doi@[]}}
\def\mn@doi@[#1]#2{\def\@tempa{#1}\ifx\@tempa\@empty \href
  {http://dx.doi.org/#2} {doi:#2}\else \href {http://dx.doi.org/#2} {#1}\fi
  \endgroup}
\def\mn@eprint#1#2{\mn@eprint@#1:#2::\@nil}
\def\mn@eprint@arXiv#1{\href {http://arxiv.org/abs/#1} {{\tt arXiv:#1}}}
\def\mn@eprint@dblp#1{\href {http://dblp.uni-trier.de/rec/bibtex/#1.xml}
  {dblp:#1}}
\def\mn@eprint@#1:#2:#3:#4\@nil{\def\@tempa {#1}\def\@tempb {#2}\def\@tempc
  {#3}\ifx \@tempc \@empty \let \@tempc \@tempb \let \@tempb \@tempa \fi \ifx
  \@tempb \@empty \def\@tempb {arXiv}\fi \@ifundefined
  {mn@eprint@\@tempb}{\@tempb:\@tempc}{\expandafter \expandafter \csname
  mn@eprint@\@tempb\endcsname \expandafter{\@tempc}}}

\bibitem[\protect\citeauthoryear{{Bhat}, {Ord}, {Tremblay}, {McSweeney}  \&
  {Tingay}}{{Bhat} et~al.}{2016}]{2016ApJ...818...86B}
{Bhat} N.~D.~R.,  {Ord} S.~M.,  {Tremblay} S.~E.,  {McSweeney} S.~J.,
  {Tingay} S.~J.,  2016, \mn@doi [\apj] {10.3847/0004-637X/818/1/86}, \href
  {https://ui.adsabs.harvard.edu/abs/2016ApJ...818...86B} {818, 86}

\bibitem[\protect\citeauthoryear{{Brisken}, {Macquart}, {Gao}, {Rickett},
  {Coles}, {Deller}, {Tingay}  \& {West}}{{Brisken}
  et~al.}{2010}]{2010ApJ...708..232B}
{Brisken} W.~F.,  {Macquart} J.~P.,  {Gao} J.~J.,  {Rickett} B.~J.,  {Coles}
  W.~A.,  {Deller} A.~T.,  {Tingay} S.~J.,   {West} C.~J.,  2010, \mn@doi
  [\apj] {10.1088/0004-637X/708/1/232}, \href
  {https://ui.adsabs.harvard.edu/abs/2010ApJ...708..232B} {708, 232}

\bibitem[\protect\citeauthoryear{{Cordes}, {Rickett}, {Stinebring}  \&
  {Coles}}{{Cordes} et~al.}{2006}]{2006ApJ...637..346C}
{Cordes} J.~M.,  {Rickett} B.~J.,  {Stinebring} D.~R.,   {Coles} W.~A.,  2006,
  \mn@doi [\apj] {10.1086/498332}, \href
  {https://ui.adsabs.harvard.edu/abs/2006ApJ...637..346C} {637, 346}

\bibitem[\protect\citeauthoryear{{Fadeev}, {Andrianov}, {Burgin}, {Popov},
  {Rudnitskiy}, {Shishov}, {Smirnova}  \& {Zuga}}{{Fadeev}
  et~al.}{2018}]{2018MNRAS.480.4199F}
{Fadeev} E.~N.,  {Andrianov} A.~S.,  {Burgin} M.~S.,  {Popov} M.~V.,
  {Rudnitskiy} A.~G.,  {Shishov} V.~I.,  {Smirnova} T.~V.,   {Zuga} V.~A.,
  2018, \mn@doi [\mnras] {10.1093/mnras/sty2055}, \href
  {https://ui.adsabs.harvard.edu/abs/2018MNRAS.480.4199F} {480, 4199}

\bibitem[\protect\citeauthoryear{{Fallows} et~al.,}{{Fallows}
  et~al.}{2014}]{2014JGRA..11910544F}
{Fallows} R.~A.,  et~al., 2014, \mn@doi [Journal of Geophysical Research (Space
  Physics)] {10.1002/2014JA020406}, \href
  {https://ui.adsabs.harvard.edu/abs/2014JGRA..11910544F} {119, 10,544}

\bibitem[\protect\citeauthoryear{{Gwinn}}{{Gwinn}}{2019}]{2019MNRAS.486.2809G}
{Gwinn} C.~R.,  2019, \mn@doi [\mnras] {10.1093/mnras/stz894}, \href
  {https://ui.adsabs.harvard.edu/abs/2019MNRAS.486.2809G} {486, 2809}

\bibitem[\protect\citeauthoryear{{Gwinn} \& {Sosenko}}{{Gwinn} \&
  {Sosenko}}{2019}]{2019MNRAS.489.3692G}
{Gwinn} C.~R.,  {Sosenko} E.~B.,  2019, \mn@doi [\mnras]
  {10.1093/mnras/stz2364}, \href
  {https://ui.adsabs.harvard.edu/abs/2019MNRAS.489.3692G} {489, 3692}

\bibitem[\protect\citeauthoryear{{Gwinn}, {Britton}, {Reynolds}, {Jauncey},
  {King}, {McCulloch}, {Lovell}  \& {Preston}}{{Gwinn}
  et~al.}{1998}]{1998ApJ...505..928G}
{Gwinn} C.~R.,  {Britton} M.~C.,  {Reynolds} J.~E.,  {Jauncey} D.~L.,  {King}
  E.~A.,  {McCulloch} P.~M.,  {Lovell} J.~E.~J.,   {Preston} R.~A.,  1998,
  \mn@doi [\apj] {10.1086/306178}, \href
  {https://ui.adsabs.harvard.edu/abs/1998ApJ...505..928G} {505, 928}

\bibitem[\protect\citeauthoryear{{Gwinn}, {Johnson}, {Smirnova}  \&
  {Stinebring}}{{Gwinn} et~al.}{2011}]{2011ApJ...733...52G}
{Gwinn} C.~R.,  {Johnson} M.~D.,  {Smirnova} T.~V.,   {Stinebring} D.~R.,
  2011, \mn@doi [\apj] {10.1088/0004-637X/733/1/52}, \href
  {https://ui.adsabs.harvard.edu/abs/2011ApJ...733...52G} {733, 52}

\bibitem[\protect\citeauthoryear{{Gwinn} et~al.,}{{Gwinn}
  et~al.}{2012}]{2012ApJ...758....6G}
{Gwinn} C.~R.,  et~al., 2012, \mn@doi [\apj] {10.1088/0004-637X/758/1/6}, \href
  {https://ui.adsabs.harvard.edu/abs/2012ApJ...758....6G} {758, 6}

\bibitem[\protect\citeauthoryear{{Hill}, {Stinebring}, {Asplund}, {Berwick},
  {Everett}  \& {Hinkel}}{{Hill} et~al.}{2005}]{2005ApJ...619L.171H}
{Hill} A.~S.,  {Stinebring} D.~R.,  {Asplund} C.~T.,  {Berwick} D.~E.,
  {Everett} W.~B.,   {Hinkel} N.~R.,  2005, \mn@doi [\apjl] {10.1086/428347},
  \href {https://ui.adsabs.harvard.edu/abs/2005ApJ...619L.171H} {619, L171}

\bibitem[\protect\citeauthoryear{{Levin} et~al.,}{{Levin}
  et~al.}{2016}]{2016ApJ...818..166L}
{Levin} L.,  et~al., 2016, \mn@doi [\apj] {10.3847/0004-637X/818/2/166}, \href
  {https://ui.adsabs.harvard.edu/abs/2016ApJ...818..166L} {818, 166}

\bibitem[\protect\citeauthoryear{{Liu}, {Pen}, {Macquart}, {Brisken}  \&
  {Deller}}{{Liu} et~al.}{2016}]{2016MNRAS.458.1289L}
{Liu} S.,  {Pen} U.-L.,  {Macquart} J.~P.,  {Brisken} W.,   {Deller} A.,  2016,
  \mn@doi [\mnras] {10.1093/mnras/stw314}, \href
  {https://ui.adsabs.harvard.edu/abs/2016MNRAS.458.1289L} {458, 1289}

\bibitem[\protect\citeauthoryear{{Pen} \& {Levin}}{{Pen} \&
  {Levin}}{2014}]{2014MNRAS.442.3338P}
{Pen} U.-L.,  {Levin} Y.,  2014, \mn@doi [\mnras] {10.1093/mnras/stu1020},
  \href {https://ui.adsabs.harvard.edu/abs/2014MNRAS.442.3338P} {442, 3338}

\bibitem[\protect\citeauthoryear{{Pen}, {Macquart}, {Deller}  \&
  {Brisken}}{{Pen} et~al.}{2014}]{2014MNRAS.440L..36P}
{Pen} U.~L.,  {Macquart} J.~P.,  {Deller} A.~T.,   {Brisken} W.,  2014, \mn@doi
  [\mnras] {10.1093/mnrasl/slu010}, \href
  {https://ui.adsabs.harvard.edu/abs/2014MNRAS.440L..36P} {440, L36}

\bibitem[\protect\citeauthoryear{{Rickett}, {Lyne}  \& {Gupta}}{{Rickett}
  et~al.}{1997}]{1997MNRAS.287..739R}
{Rickett} B.~J.,  {Lyne} A.~G.,   {Gupta} Y.,  1997, \mn@doi [\mnras]
  {10.1093/mnras/287.4.739}, \href
  {https://ui.adsabs.harvard.edu/abs/1997MNRAS.287..739R} {287, 739}

\bibitem[\protect\citeauthoryear{{Rickett}, {Stinebring}, {Coles}  \&
  {Jian-Jian}}{{Rickett} et~al.}{2011}]{2011AIPC.1357...97R}
{Rickett} B.,  {Stinebring} D.,  {Coles} B.,   {Jian-Jian} G.,  2011, in
  {Burgay} M.,  {D'Amico} N.,  {Esposito} P.,  {Pellizzoni} A.,   {Possenti}
  A.,  eds,  American Institute of Physics Conference Series Vol. 1357,
  American Institute of Physics Conference Series. pp 97--100,
  \mn@doi{10.1063/1.3615088}

\bibitem[\protect\citeauthoryear{{Simard} \& {Pen}}{{Simard} \&
  {Pen}}{2018}]{2018MNRAS.478..983S}
{Simard} D.,  {Pen} U.-L.,  2018, \mn@doi [\mnras] {10.1093/mnras/sty1140},
  \href {https://ui.adsabs.harvard.edu/abs/2018MNRAS.478..983S} {478, 983}

\bibitem[\protect\citeauthoryear{{Simard}, {Pen}, {Marthi}  \&
  {Brisken}}{{Simard} et~al.}{2019a}]{2019MNRAS.488.4952S}
{Simard} D.,  {Pen} U.~L.,  {Marthi} V.~R.,   {Brisken} W.,  2019a, \mn@doi
  [\mnras] {10.1093/mnras/stz2043}, \href
  {https://ui.adsabs.harvard.edu/abs/2019MNRAS.488.4952S} {488, 4952}

\bibitem[\protect\citeauthoryear{{Simard}, {Pen}, {Marthi}  \&
  {Brisken}}{{Simard} et~al.}{2019b}]{2019MNRAS.488.4963S}
{Simard} D.,  {Pen} U.~L.,  {Marthi} V.~R.,   {Brisken} W.,  2019b, \mn@doi
  [\mnras] {10.1093/mnras/stz2046}, \href
  {https://ui.adsabs.harvard.edu/abs/2019MNRAS.488.4963S} {488, 4963}

\bibitem[\protect\citeauthoryear{{Stinebring}, {McLaughlin}, {Cordes},
  {Becker}, {Goodman}, {Kramer}, {Sheckard}  \& {Smith}}{{Stinebring}
  et~al.}{2001}]{2001ApJ...549L..97S}
{Stinebring} D.~R.,  {McLaughlin} M.~A.,  {Cordes} J.~M.,  {Becker} K.~M.,
  {Goodman} J.~E.~E.,  {Kramer} M.~A.,  {Sheckard} J.~L.,   {Smith} C.~T.,
  2001, \mn@doi [\apjl] {10.1086/319133}, \href
  {https://ui.adsabs.harvard.edu/abs/2001ApJ...549L..97S} {549, L97}

\bibitem[\protect\citeauthoryear{{Stinebring}, {Hill}, {McLaughlin}, {Becker},
  {Cordes}  \& {Kramer}}{{Stinebring} et~al.}{2003}]{2003ASPC..302..263S}
{Stinebring} D.~R.,  {Hill} A.~S.,  {McLaughlin} M.~A.,  {Becker} K.~M.,
  {Cordes} J.~M.,   {Kramer} M.,  2003, in {Bailes} M.,  {Nice} D.~J.,
  {Thorsett} S.~E.,  eds,  Astronomical Society of the Pacific Conference
  Series Vol. 302, Radio Pulsars. p.~263

\bibitem[\protect\citeauthoryear{{Tuntsov}, {Bignall}  \& {Walker}}{{Tuntsov}
  et~al.}{2013}]{2013MNRAS.429.2562T}
{Tuntsov} A.~V.,  {Bignall} H.~E.,   {Walker} M.~A.,  2013, \mn@doi [\mnras]
  {10.1093/mnras/sts527}, \href
  {https://ui.adsabs.harvard.edu/abs/2013MNRAS.429.2562T} {429, 2562}

\bibitem[\protect\citeauthoryear{{Walker} \& {Stinebring}}{{Walker} \&
  {Stinebring}}{2005}]{2005MNRAS.362.1279W}
{Walker} M.~A.,  {Stinebring} D.~R.,  2005, \mn@doi [\mnras]
  {10.1111/j.1365-2966.2005.09396.x}, \href
  {https://ui.adsabs.harvard.edu/abs/2005MNRAS.362.1279W} {362, 1279}

\bibitem[\protect\citeauthoryear{{Walker}, {Melrose}, {Stinebring}  \&
  {Zhang}}{{Walker} et~al.}{2004}]{2004MNRAS.354...43W}
{Walker} M.~A.,  {Melrose} D.~B.,  {Stinebring} D.~R.,   {Zhang} C.~M.,  2004,
  \mn@doi [\mnras] {10.1111/j.1365-2966.2004.08159.x}, \href
  {https://ui.adsabs.harvard.edu/abs/2004MNRAS.354...43W} {354, 43}

\bibitem[\protect\citeauthoryear{{Walker}, {Koopmans}, {Stinebring}  \& {van
  Straten}}{{Walker} et~al.}{2008}]{2008MNRAS.388.1214W}
{Walker} M.~A.,  {Koopmans} L.~V.~E.,  {Stinebring} D.~R.,   {van Straten} W.,
  2008, \mn@doi [\mnras] {10.1111/j.1365-2966.2008.13452.x}, \href
  {https://ui.adsabs.harvard.edu/abs/2008MNRAS.388.1214W} {388, 1214}

\bibitem[\protect\citeauthoryear{{Xu} et~al.,}{{Xu}
  et~al.}{2018}]{2018MNRAS.476.5579X}
{Xu} Y.~H.,  et~al., 2018, \mn@doi [\mnras] {10.1093/mnras/sty566}, \href
  {https://ui.adsabs.harvard.edu/abs/2018MNRAS.476.5579X} {476, 5579}

\makeatother
\end{thebibliography}



\appendix




\bsp	
\label{lastpage}
\end{document}